\RequirePackage{fix-cm}
\documentclass[smallextended]{svjour3}       
\smartqed  
\usepackage{graphicx}
\usepackage{amsmath}
\usepackage[utf8]{inputenc} 
\usepackage{natbib}
\usepackage{siunitx}
\usepackage{bm}
\usepackage{mathabx} 
\usepackage{dsfont}  
\usepackage{booktabs}

\usepackage[%
breaklinks=true,%
colorlinks=true,%
citecolor=blue,%
pdfauthor={D. Amato et al.},%
pdftitle={Non-averaged regularized formulations as an alternative to semi-analytical orbit propagation methods}%
]{hyperref}

\makeatletter
\def\cl@chapter{\@elt {theorem}}
\makeatother
\usepackage{cleveref}

\DeclareSIUnit\SFU{SFU}
\newcommand{\STELA}{\texttt{STELA}{}}
\newcommand{\Thalassa}{\texttt{THALASSA}{}}
\newcommand{\citeurl}[2]{\footnote{\textsc{URL:} \url{#1}, last visited: #2.}}

\newcommand{\dt}[1]{\frac{\mathrm{d}#1}{\mathrm{d}t}}
\newcommand{\distR}{\mathcal{R}}
\newcommand{\distEarth}{\mathcal{R}_\Earth}
\newcommand{\distPert}{\mathcal{R}'}
\newcommand{\oscE}{\hat{E}}
\newcommand{\oscRefE}{\hat{E}^\mathrm{R}}
\newcommand{\LSODAR}{\texttt{LSODAR}}

\crefname{equation}{equation}{equations}
\Crefname{equation}{Equation}{Equations}
\crefname{figure}{figure}{figures}
\Crefname{figure}{Figure}{Figures}

\journalname{Celestial Mechanics and Dynamical Astronomy}
\begin{document}

\title{Non-averaged regularized formulations as an alternative to semi-analytical orbit propagation methods
\thanks{This work is partially funded by the European Commission's Framework Programme 7, through the Stardust Marie Curie Initial Training Network, FP7-PEOPLE-2012-ITN, Grant Agreement 317185.}
}
%

\titlerunning{Non-averaged regularized formulations}        

\author{Davide Amato        \and
        Claudio Bombardelli \and
        Giulio Baù          \and
        Vincent Morand      \and
        Aaron J. Rosengren 
}


\institute{D. Amato \at
           School of Aerospace Engineering, Technical University of Madrid, Madrid, Spain. \\
           \email{davideamato@email.arizona.edu}             \\
           \emph{Present address:} Department of Aerospace and Mechanical Engineering, The University of Arizona, Tucson, AZ, United States.
           \and
           C. Bombardelli \at
           School of Aerospace Engineering, Technical University of Madrid, Madrid, Spain.
           \and
           G. Baù \at
           Mathematics Department, University of Pisa, Pisa, Italy.
           \and
           V. Morand \at
           Centre National d'Études Spatiales, Toulouse, France.
           \and
           A. J. Rosengren \at
           Department of Aerospace and Mechanical Engineering, The University of Arizona, Tucson, AZ, United States.
}

\date{Received: date / Accepted: date}

\maketitle

\begin{abstract}
This paper is concerned with the comparison of semi-analytical and non-averaged propagation methods for Earth satellite orbits.
We analyse the total integration error for semi-analytical methods and {propose a novel decomposition} into dynamical, model truncation, short-periodic, and numerical error components.
The first three are attributable to distinct approximations required by the method of averaging, which fundamentally limit the attainable accuracy.
In contrast, numerical error, the only component present in non-averaged methods, can be significantly mitigated by employing adaptive numerical algorithms and regularized formulations of the equations of motion.
We present a collection of non-averaged methods based on the integration of {existing} regularized formulations of the equations of motion through an adaptive solver.
{We implemented the collection in the orbit propagation code \Thalassa{}, which we make publicly available, and we compared the non-averaged methods} to the semi-analytical method implemented in the orbit propagation tool \STELA{} through numerical tests involving long-term propagations (on the order of decades) of LEO, GTO, and high-altitude HEO orbits.
For the test cases considered, regularized non-averaged methods were found to be up to two times slower than semi-analytical for the LEO orbit, to have comparable speed for the GTO, and to be ten times as fast for the HEO (for the same accuracy).
{We show for the first time that} efficient implementations of non-averaged regularized formulations of the equations of motion, and especially of non-singular element methods, are attractive candidates for the long-term study of high-altitude and highly elliptical Earth satellite orbits.
\keywords{numerical methods \and regularization \and special perturbations \and semi-analytical methods}
\end{abstract}


\section{Introduction}
Predicting the evolution of the population of Earth satellites requires fast orbit propagation techniques that are capable of efficiently taking into account a plethora of physical phenomena and dynamical configurations.
A comprehensive picture of the evolution of the near-Earth environment is usually constructed through large-scale Monte Carlo simulations, which are computationally burdensome.
These simulations are often performed either through general perturbations methods, which rely on approximate analytical solutions of the perturbed gravitational two-body problem, or through semi-analytical methods, in which averaged equations of motion are integrated numerically.

Due to their decisive speed advantage, general perturbations theories are employed to propagate large ensembles of objects, especially when frequent orbital updates are available or when only statistical results are required, with relaxed accuracy requirements for individual objects.
Perhaps the most widely used general perturbations theory is SGP4, a simplified version of the SGP (Simplified General Perturbations) theory.
SGP4 was originally used to propagate the Two-Line-Element (TLE) catalog produced and maintained by the US Joint Space Operations Center~\citep{Vallado2006}.
SGP4 is based on the Brouwer analytical solution of the main satellite problem that includes zonal terms of the Earth's gravity potential up to the fifth order, and drag computed through a power law for the atmospheric density~\citep{Hoots2004}.
The Debris Cloud Propagator (DCP) included in the SDM suite~\citep{Rossi2009} is a general perturbations software used to study the evolution of the population of orbital debris.
DCP adopts a faster and less sophisticated approach than SGP4, using analytical solutions for the semi-major axis, eccentricity, and inclination under the effect of atmospheric drag, and for the longitude of node and argument of perigee under the effect of the Earth's oblateness~\citep{Rossi2009,Anselmo1997}.
Recently, \citet{Moeckel2015} introduced the analytical propagation code Ikebana, a parallelized version of the FLORA orbit propagator.
Ikebana makes use of modern parallel computing capabilities endowed by GPUs to accelerate the propagation of large object populations for applications to debris models, and uses single precision wherever possible to increase the performance of GPU calculations.
Both FLORA and Ikebana were validated against a non-averaged code and by reproducing the overall long-term behavior of historic TLE data for the Vanguard-1 satellite.
{Short-term analytical solutions (in the order of a few revolutions) using the Kustaanheimo-Stiefel regularization~\citep{KS1965} have also been derived, for instance by including perturbations from zonal harmonics~\citep[and references therein]{Sellamuthu2018} and lunar gravitation~\citep{Sellamuthu2017}.}


Semi-analytical techniques offer a compromise between general and special perturbations.
In these approaches, the fast variables are eliminated from the equations through the \emph{method of averaging}, which consists in the elimination of high-frequency components from the equations of motion by an analytical or numerical averaging over a short time scale.
If averaging is correctly performed, the time scale of the problem changes from one characteristic of the orbital period to that of the long-periodic effects, thus enabling a numerical solver to take larger step sizes with respect to the osculating orbit.
In essence, semi-analytical techniques employ information on the approximate analytical solutions provided by general perturbations to improve the efficiency of special perturbations, at the cost of accuracy.
This process may permit significant savings in computational time with respect to the integration of the equations of motion in Cartesian coordinates, i.e., the Cowell formulation.

Semi-analytical propagation algorithms are widely employed for the future prediction of the circumterrestrial debris environment.
The European Space Agency MASTER debris model uses the semi-analytical orbit propagation code FOCUS (Fast Orbit Computation Utility Software), which integrates single-averaged Gauss equations for equinoctial elements.
A faster propagation code denominated DELTOP is also used for the prediction of the future evolution of the MASTER model. 
DELTOP employs a simple Euler integration scheme to decrease computational time~\citep[chapter 5]{Klinkrad2006}.
The software was validated against accurate solutions, suggesting that the employment of mean Keplerian elements compensates the inaccuracy of the Euler integration scheme~\citep[chapter 1]{Dahlquist1974}.
NASA uses the semi-analytical propagators PROP3D and GEOPROP to update its LEGEND debris population model~\citep{Liou2004}.
PROP3D is employed for all the orbital regimes except GEO.
It includes perturbations from atmospheric drag, zonal harmonics of the geopotential up to $J_4$, lunisolar gravitational accelerations, and solar radiation pressure.
GEOPROP is used for the GEO population and is based on the averaged approach by~\citet{vanderHa1986}.
The semi-analytical propagator FOP, which is included in the SDM suite by ~\citet{Rossi2009}, is an optimized version of the LOP code developed by~\citet{Kwok1986}.

Orbit propagators used for Space Situational Awareness (SSA) are, in general, more sophisticated than those used for debris modeling.
The Draper Semi-analytic Satellite Theory (DSST) was one of the first propagators employed in SSA, and it includes all the principal perturbations affecting Earth satellite orbits (gravitational perturbations from the Earth's zonal and tesseral harmonics, the Sun, and the Moon, and non-gravitational perturbations from atmospheric drag and solar radiation pressure) in the equations of motion for mean equinoctial elements.
Besides, it retains long-periodic and secular trends arising from tesseral resonances and it includes higher-order cross-coupling terms between the geopotential and atmospheric drag.
Although there is no principal reference for the DSST, a comprehensive summary of the theory and equations has been given by~\citet{Danielson1995}.
\citet{Golikov2012} presented the numerical code THEONA, based on an elegant approach in which the equations of motion describe the deviations from an \emph{intermediate} orbit corresponding to the exact solution of the generalized problem of two fixed centers.
THEONA has been used for orbit prediction in Soyuz and Progress missions.
The French space agency CNES (Centre Nationale d'Études Spatiales), developed the semi-analytical propagator \STELA{} (Semi-Analytical Tool for End-of-Life Analysis) to verify the compliance of spacecraft with the end-of-life regulations detailed in the French Space Operations Act~\citep{LeFevre2014}.
\STELA{} integrates the equations of motion for mean equinoctial elements including all the main perturbations acting on Earth satellite orbits.
The averaging theory is carried out at first order for all perturbations except $J_2$, for which it is carried out at second order.
Cross-coupling between the oblateness and atmospheric drag perturbations is also considered.
Short-periodic terms can be recovered for the $J_2$ and lunisolar perturbations.
Due to its excellent state of validation and its public availability, we use \STELA{} as the reference semi-analytical propagator against which to compare the code that we introduce in this paper.

A thorough understanding of the dynamical characteristics of the system is required in order to obtain reliable results and optimize the performance of semi-analytical methods.
If two or more angular frequencies are commensurable, the corresponding terms of the perturbing function change according to a long-periodic or secular behavior.
This situation is ubiquitous in celestial mechanics, and is {known as a state of} \emph{resonance}~\citep{Murray1999}.
For instance, the navigational satellites in MEO and the telecommunications satellites in GEO are in resonance with the tesseral harmonics of the Earth's geopotential to satisfy {ground-track repeatability} requirements.
Secular and semi-secular resonances with the Sun and the Moon can be exploited to drastically reduce orbit lifetimes~\citep{Gkolias2016}, while lunar mean-motion resonances are being exploited to achieve stable highly elliptical orbits~\citep{Dichmann2013}.
Specific provisions have to be adopted during the averaging process in the presence of resonances, otherwise long-periodic and secular trends in the orbital elements will be missed, resulting in an incorrect trajectory.
In the case of tesseral resonances, some of the terms depending on the mean anomaly (that would otherwise be ``averaged out'') have to be retained in the geopotential~\citep{Kaula1966}, and similar techniques are used for other types of resonances~\citep{Morbidelli2002}.
High-order averaging schemes are required to capture coupling effects between different perturbations.
This is exemplified by the dynamics of GTOs, in which the coupling between $J_2$, solar {gravity}, and drag perturbations makes the trajectories extremely sensitive to the initial conditions and to uncertainty in the state and in physical parameters~\citep{Lamy2011}.\footnote{Regardless of the orbit propagation method, uncertainties in the orbit determination, in the predictions of solar activity, and in the modeling of the atmosphere-spacecraft interaction make GTOs unpredictable in the long term.}
Even if these dynamical configurations are duly accounted for, semi-analytical techniques are intrinsically limited in their accuracy due to the approximations introduced in the averaging process, and some parameters, such as the orders of truncation of the perturbing functions, have to be judiciously chosen before integrating the rates of change of the mean elements.
Moreover, averaging over the fast variables necessarily involves the loss of information on the short-periodic variations in the orbital elements.
The information can only be partially recovered by adding short-periodic terms (which are computed analytically) to the mean elements.
As we will show in this paper, this process introduces errors in the calculation of the osculating elements from the mean.
On the other hand, semi-analytical methods may be faster by up to two orders of magnitude with respect to the Cowell formulation~\citep{Setty2016}.

In contrast to semi-analytical techniques, the accuracy of special perturbations is only limited by the physical model and by the available processing power and memory. Once the physical model is defined, the solver can be easily tuned by changing the integration time step.
Regardless of these advantages, special perturbations (or \emph{non-averaged}) methods have not found widespread use in the integration of large sets of initial conditions until recently, probably due to the lack of the necessary computational resources.
However, these circumstances are changing.

Already in 1997, \citet{Coffey1998} demonstrated that maintaining the US space object catalog\footnote{The US space object catalog is a public database of orbital data for more than \num{17000} objects (at the time of writing). Access to the catalog is available at \url{https://www.space-track.org/}, last visited: July 6\textsuperscript{th}, 2018.} through special perturbations could easily be achieved, given enough computational hardware.
In fact today, this catalog is fully maintained by the US Joint Space Operations Center using special perturbations.
A wide range of single- and multi-step numerical solvers for the Cowell formulation has been implemented in the GMAT\citeurl{http://gmatcentral.org/}{October 8\textsuperscript{th}, 2018} and Copernicus~\citep{Williams2010} software suites.

There are some drawbacks in the Cowell formulation.
First, the solution of the equations of motion in Cartesian coordinates is unstable in the Lyapunov sense, even in the unperturbed problem~\citep{Bond1996}.
Then, since the gravitational potential is singular at collision, the right-hand side of the equations of motion exhibit strong oscillations when the particle is close to the main body.
Furthermore, because the {physical time is used as the independent variable, the distribution of steps along the orbit is shifted towards the apoapsis for constant step size, a fact that is particularly detrimental for highly eccentric orbits.}
These disadvantages can be mitigated by resorting to \emph{regularized formulations}.


A \emph{regularization} of the two-body problem is an analytical procedure that removes the singularity of collision from the vector field. It usually consists of three steps: introducing a new independent variable instead of time (which is called \emph{fictitious time}), transforming the Cartesian coordinates of position and velocity into new variables, and embedding the integrals of the motion into the transformed equations. As a result the new differential equations are linear with constant coefficients when the motion is unperturbed. Additionally, depending upon the transformation of time and the spatial variables, the solution can be analytically continued through the collision. This is the case for the regularizations due to Moser~\citep[][section 1.6]{MoserZehnder2005},~\citet{Sperling1961}, and~\citet{KS1965}. Other regularizations, like the one developed by~\citet{Burdet1969} and~\citet{Ferrandiz1988}, lead to formulations that still enjoy the linear form of the transformed system, but are not defined at collision.
Regularized formulations are superior in terms of accuracy and computational cost than Cowell's method.
However, additional operations are required to obtain the position and velocity at a prescribed epoch starting from the new spatial variables and the fictitious time{, which could be the reason why they are not commonly implemented in orbit propagators.}

Since regularizations transform the nonlinear Newtonian equations of the two-body problem into a set of linear equations, variation of parameters (VOP) methods can be easily developed from them. These formulations use a set of orbital elements, i.e., quantities that are constant along the unperturbed solution, to propagate the state also when perturbations are applied. The new elements inherit the advantageous properties of their parent variables, for example they may be regular at collision, with the advantage over the latter that if perturbations are weak they can reach the same accuracy with much fewer integration steps. This feature is particularly attractive because Earth satellite orbits can often be regarded as weakly perturbed.\footnote{Exceptions are given by translunar orbits, impulsive maneuvers, and the terminal phase of re-entry trajectories.}
Recently, \cite{Bau2015} have proposed a VOP formulation, here referred to as EDromo, which appears to be particularly promising for efficient orbit propagation. It was noticed that, especially for highly eccentric orbits, this method shows an excellent performance when compared to many other regularized schemes.
The reader is addressed to the monograph by~\citet{Roa2017} for a comprehensive overview of regularization theory and its applications.

In this paper, we show that special perturbation methods based on regularized formulations can compete and even perform better than semi-analytical methods for the long-term propagations (on the order of decades) of objects orbiting around the Earth.
Note that for this kind of applications the Cowell formulation is never used because of the small step sizes required, which cause strong accumulation of round-off error and long computational times.
In order to carry out {this} study we developed a Fortran code, named \Thalassa{}, which includes Cowell's method, EDromo, the Kustaanheimo Stiefel (KS) regularization~\citep{KS1965}, and a set of regular elements that were obtained by \citet[][section 19]{Stiefel1971} from KS.
A sophisticated numerical solver, named \LSODAR{} (Livermore Solver for Ordinary Differential equations with Automatic Root-finding), has been included to integrate the differential equations of motion.
Some results using a preliminary version of \Thalassa{} for cis- and translunar orbits have been shown in \citet{Amato2018}.

\Thalassa{} is compared to the \STELA{} orbit propagator through numerical experiments performed for a Low Earth Orbit (LEO), a Geostationary Transfer Orbit (GTO), and a high-altitude, Highly Elliptical Orbit (HEO).
The test cases have been chosen as to maximize the scientific interest and the intrinsic difficulty in obtaining accurate position and velocity on decadal timescales.
Symplectic integration methods, which are based on the rigorous conservation of an approximate Hamiltonian of the problem, are commonly used in astrophysical research to perform extremely long integrations and may seem like a feasible candidate for the study.
Nevertheless, we do not take them into account in this work since previous research has shown that the conservation of the symplectic structure does not necessarily imply the reduction of errors in position and velocity~\citep{Amato2017}.

The paper is organized as follows. In \cref{sec:avg_method} we provide an overview of the method of averaging. In \cref{sec:avg_err_analysis} we present an analytical breakdown of the errors arising in the integration of averaged equations of motion, which is also an original contribution of the paper.
\STELA{} and \Thalassa{}, the two codes used in the study, are described in \cref{sec:software}, and the numerical results are presented in \crefrange{sec:LEO_test}{sec:HEO_test}. \Cref{sec:conclu} contains the conclusions.
  
\section{Method of averaging}
\label{sec:avg_method}
In this section, we summarize the theory of averaging in equinoctial elements as presented by~\cite{Danielson1995}, using the expressions for the perturbing functions presented by~\cite{Giacaglia1977}.
This set of elements underlies many of the most widely known semi-analytical propagators, such as the Semi-analytical Tool for End-of-life Analysis (\STELA{}) and the DSST.

\subsection{Osculating equations of motion and perturbing functions}
We show the osculating (i.e., non-averaged) equations of motion for the set of equinoctial elements $\bm{E}$ as derived by \citet{Giacaglia1977} and \citet{Nacozy1977}, which is expressed in terms of the classical orbital elements as
\begin{equation}
\renewcommand\arraystretch{1.8}
\bm{E} = \begin{Bmatrix} a \\ h \\ k \\ P \\ Q \\ \lambda \end{Bmatrix} =
\begin{Bmatrix} a \\
 e \sin \left( \omega + \Omega \right) \\
 e \cos \left( \omega + \Omega \right) \\
 \sin \dfrac{i}{2} \cos \Omega \\[6pt]
 \sin \dfrac{i}{2} \sin \Omega \\
 M + \omega + \Omega
\end{Bmatrix}.
\label{eq:equ_els}
\end{equation}
This set only presents a singularity for $i = \SI{180}{\degree}$, a case that is of limited interest for  applications to Earth satellites.
Letting $\gamma = \sqrt{1 - h^2 - k^2} = \sqrt{1 - e^2}$, the rates of change of the osculating elements are
\begin{equation}
\begin{aligned}
\dt{a} &= \dfrac{\strut 2}{\strut na} \distR_\lambda \\
\dt{h} &= - \dfrac{\gamma}{na^2 (1 + \gamma)} h \distR_\lambda + \dfrac{\gamma}{na^2} \distR_k + \dfrac{1}{2na^2 \gamma} k \left( P \distR_P + Q \distR_Q \right) \\
\dt{k} &= - \dfrac{\gamma}{na^2 (1 + \gamma)} k \distR_\lambda - \dfrac{\gamma}{na^2} \distR_h - \dfrac{1}{\strut 2na^2 \gamma} h \left( P \distR_P + Q \distR_Q \right) \\
\dt{P} &= -\dfrac{1}{2na^2 \gamma} P \distR_\lambda - \dfrac{1}{4na^2 \gamma} \distR_Q + \dfrac{\strut 1}{\strut 2na^2 \gamma} P \left( h \distR_k - k \distR_h \right) \\
\dt{Q} &= -\dfrac{1}{2na^2 \gamma} Q \distR_\lambda + \dfrac{1}{4na^2 \gamma} \distR_P + \dfrac{ 1}{ 2na^2 \gamma} Q \left( h \distR_k - k \distR_h \right) \\
\dt{\lambda} &= n - \dfrac{\strut 2}{na} R_a + \dfrac{\gamma}{2na^2} \left(k \distR_k + h \distR_h \right) + \dfrac{1}{2na^2 \gamma} \left( P \distR_P + Q \distR_Q \right)
\end{aligned}.
\label{eq:osc_EoMs_Giacaglia}
\end{equation}
Denoting with $E_i \: (i = 1, \ldots, 6)$ a generic equinoctial element, we take into account both conservative and dissipative perturbations in the quantity $\distR_{E_i}$:
\[
\distR_{E_i} = \frac{\partial \distR}{\partial E_i} + \bm{q} \cdot \frac{\partial \bm{r}}{\partial E_i},
\]
where $\bm{q}$ is the vector sum of the dissipative perturbing accelerations. In the perturbing function $\distR$, we consider perturbations due to the non-spherical gravity field of the Earth and to the Moon and the Sun considered as point masses,
\[
\distR = \distEarth + \distR_\Sun + \distR_\leftmoon.
\]
According to \citet[p. 31]{Kaula1966}, the perturbing function $\distEarth$ is expanded in spherical coordinates as
\begin{equation}
\distEarth = \sum_{l=1}^{\infty} \sum_{m=0}^{l} \frac{\mu_\Earth a^l_\Earth}{r^{l+1}} P_{lm} \left( \sin \phi \right) \left( C_{lm} \cos{m}L + S_{lm}\sin{m}L \right),
\end{equation}
where $a_\Earth$ is the mean equatorial radius of the Earth, $P_{lm}$ are the associated Legendre functions of order $l$ and degree $m$, $\phi$ and $L$ are respectively the geographic latitude and longitude, and $C_{lm}$, $S_{lm}$ are coefficients that are determined empirically.
For an axially symmetric body $C_{11} = S_{11} = 0$, therefore we will consider the outer sum to always start from $l = 2$ in the following.
The expression for $\distEarth$ in equinoctial elements is
\begin{equation}
\distEarth = \sum_{l=2}^{\infty} \sum_{m=0}^{l} \sum_{p=0}^{l} \sum_{q=-\infty}^{\infty} \mathcal{R}_{lmpq}.
\label{eq:R}
\end{equation}
Each of the terms $\mathcal{R}_{lmpq}$ is written as
\begin{equation}
\begin{split}
\mathcal{R}_{lmpq} = \frac{\mu_\Earth a^l_\Earth}{a^{l+1}} J_{lmp}(c) K_{lpq}(\gamma) \times \\ \times \left[ \mathds{R}_{lmpq}(h,k,P,Q) \left( A_{lm} \cos{\psi_{lmpq}} + B_{lm} \sin{\psi_{lmpq}} \right) + \right. \\ \left. + \mathds{I}_{lmpq}(h,k,P,Q) \left( A_{lm} \sin{\psi_{lmpq}} - B_{lm} \cos{\psi_{lmpq}} \right) \right],
\end{split}
\label{eq:Rlmpq}
\end{equation}
where
\[
A_{lm} = \left\lbrace \begin{aligned} C_{lm}, \quad &l-m \: \text{even} \\ -S_{lm}, \quad &l-m \: \text{odd}\end{aligned} \right. , \qquad
B_{lm} = \left\lbrace \begin{aligned} S_{lm}, \quad &l-m \: \text{even} \\ C_{lm}, \quad &l-m \: \text{odd}\end{aligned} \right. ,
\]
and $c = \sqrt{1 - P^2 - Q^2} = \cos \left(i/2\right)$.
The quantities $J_{lmp}$ and $K_{lpq}$ are, respectively, a polynomial in $c$ and an infinite power series in $\gamma$.
The functions $\mathds{R}_{lmpq}$ and $\mathds{I}_{lmpq}$ are finite power series of their arguments.
The angle $\psi_{lmpq}$,
\begin{equation}
\psi_{lmpq} = (l - 2p + q)\lambda - m\theta,
\label{eq:psi_Earth}
\end{equation}
is a linear combination of the mean longitude $\lambda$ and the Greenwich Mean Sidereal Time (GMST) $\theta$.

The perturbing function due to the Moon and the Sun as point masses takes the same form $\distPert$ for each of the bodies,
\begin{equation}
\distPert = \sum_{l=2}^{\infty} \sum_{m=0}^{l} \sum_{p=0}^{l} \sum_{p'=0}^{l} \sum_{q=-\infty}^{\infty} \sum_{q'=-\infty}^{\infty} \distPert_{lmpqp'q'},
\label{eq:R1}
\end{equation}
where the expression for the terms $\distPert_{lmpqp'q'}$ is the following:
\begin{equation}
\begin{split}
\distPert_{lmpqp'q'} = \mu' (n')^2 \frac{a^l}{(a')^{l-2}} \varepsilon_m \frac{\left(l-m\right)!}{\left(l+m\right)!} J_{lmp}(c) L_{lpq}(\gamma) F_{lmp'}(i') G_{lp'q'}(e') \times \\ \times \left( \mathds{R}_{lmpq} \cos \psi'_{lmpqp'q'} + \mathds{I}_{lmpq} \sin \psi'_{lmpqp'q'} \right).
\end{split}
\label{eq:R1lmpq}
\end{equation}
In \cref{eq:R1lmpq}, primed quantities pertain to the perturbing body.
In particular, $\mu'$ is its gravitational parameter, and $n'$ is its mean motion.
The eccentricity functions $L_{lpq}$ and $G_{lp'q'}$ are infinite power series, and the inclination function $F_{lmp'}$ is a polynomial in the trigonometric functions of the inclination.
The factor $\varepsilon_m$ is equal to \num{1} for $m = 0$, and is equal to \num{2} for $m \neq 0$.
The angle $\psi'_{lmpqp'q'}$ is defined as
\begin{equation}
\begin{split}
\psi'_{lmpqp'q'} = \left(l - 2p + q \right)\lambda - \left( l - 2p' + q' \right) \lambda' + q'\left( \omega' + \Omega' \right) - \\
- \left(m + 2p' - l\right)\Omega'.
\end{split}
\label{eq:psi_pert}
\end{equation}
All of the primed quantities in the right-hand side of the above equation are usually considered as functions of time that are slow with respect to the mean motion.
For instance, when considering a MEO satellite perturbed by the Moon, $\dot{\lambda}$ is in the order of \num{12} hours while $\dot{\lambda}'$ is equal to about \num{27} days.

In the rest of the paper, primed quantities will always refer to the lunisolar gravitational perturbations.

\subsection{Averaging perturbing functions}
We summarize here the core details of the method of averaging perturbations stemming from perturbing functions as presented in \citet{Danielson1995}.
Also, we denote osculating orbital elements with a hat in the following.
We define the \emph{mean} elements $\bm{E}$ through their relation with the \emph{osculating} elements $\hat{\bm{E}}$,
\begin{equation}
\hat{E}_i = E_i + \sum_{j=1}^{\infty} \epsilon^j \eta_{i,j} \left( \bm{E}, t \right),\qquad i = 1, \ldots, 6,
\label{eq:osc_mean_NI}
\end{equation}
where $\epsilon$ is a small perturbation parameter.
The terms $\epsilon^j \eta_{i,j}$ are small, short-periodic variations that are added to the mean elements $E_i$ to obtain the osculating elements $\hat{E}_i$.
They explicitly depend on time, since perturbations as tesseral harmonics of the Earth's gravity potential, and as those due to perturbing bodies, are also explicit in time.

The perturbed two-body problem is stated in osculating elements as
\begin{equation}
\frac{\mathrm{d}\hat{E}_i}{\mathrm{d}t} = n(\hat{a}) \delta_{i,6} + \epsilon F_i (\hat{\bm{E}},t),
\label{eq:osc_EoMs}
\end{equation}
where $\delta_{i,6}$ is the Kronecker delta and $n$ is the mean motion, which is a function of the osculating semi-major axis $\hat{a}$.
Note that $\epsilon F_i$ includes all the terms appearing on the right-hand side of \cref{eq:osc_EoMs_Giacaglia} that contain the perturbing forces.
Our aim is to write the averaged equations of motion in the form
\begin{equation}
\frac{\mathrm{d}{E}_i}{\mathrm{d}t} = n\left(a \right) \delta_{i,6} + \sum_{j=1}^{\infty} \epsilon^j A_{i,j} \left(a,h,k,P,Q,t\right),
\label{eq:mean_EoMs}
\end{equation}
where the right-hand side represents a power series in the perturbation parameter $\epsilon$ with the coefficients given by $A_{i,j}$.
Since the high-frequency components are relegated to the short-periodic terms, the rates of change of the mean elements are small, that is
\begin{align*}
\frac{1}{na} \left| \frac{\mathrm{d}a}{\mathrm{d}t} \right| &\ll 1, \\ 
\frac{1}{n} \left| \frac{\mathrm{d}E_i}{\mathrm{d}t} \right| &\ll 1, \quad \text{for} \: i = 2, \ldots, 5, \\ 
\left| \frac{1}{n} \frac{\mathrm{d}\lambda}{\mathrm{d}t} - 1 \right| &\ll 1.
\end{align*}
As to find a suitable form for the coefficients $A_{i,j}$ in \cref{eq:mean_EoMs}, we first express $n(\hat{a})$ and $F_i(\hat{\bm{E}},t)$ as functions of the mean elements by expanding them in power series of $\epsilon$,
\begin{align}
n\left(\hat{a} \right) &= n \left( a \right) + \sum_{j=1}^{\infty} \epsilon^j N_j \left( a \right) \label{eq:n_mean} \\
F_i ( \hat{\bm{E}}, t ) &= F_i \left(\bm{E}, t \right) + \sum_{j=1}^{\infty} \epsilon^j f_{i,j} \left( \bm{E}, t \right). \label{eq:F_mean}
\end{align}
Then, we differentiate \cref{eq:osc_mean_NI} with the use of \cref{eq:mean_EoMs} and set the resulting expressions for the derivatives equal to those in \cref{eq:osc_EoMs}, where we have used \cref{eq:n_mean,eq:F_mean}.
The \emph{equations of averaging} take the form
\begin{equation}
\sum_{j=1}^{\infty} \epsilon^j \left( A_{i,j} + \frac{\partial \eta_{i,j}}{\partial \bm{E}} \cdot \frac{\mathrm{d} \bm{E}}{\mathrm{d}t} + \frac{\partial \eta_{i,j}}{\partial t} \right) =
\epsilon F_i + \sum_{j=1}^{\infty} \epsilon^j \left( \delta_{i,6} N_j \left( a \right) + \epsilon f_{i,j} \right).
\label{eq:averaging}
\end{equation}
By finding $A_{i,j}$ and $\eta_{i,j}$ such that they satisfy \cref{eq:averaging} up to the $M$-th order, we can build an $M$-th order averaged theory.
We will limit ourselves to the first order in this study; more details on how to derive higher order theories are found in \citet{Danielson1995}.
We have
\begin{equation}
A_{i,1} + \frac{\partial \eta_{i,1}}{\partial \lambda} n\left( a \right) + \frac{\partial \eta_{i,1}}{\partial t} = F_i + \delta_{i,6} N_1 (a).
\label{eq:avg_first}
\end{equation}
We now define the \emph{single-averaging operator} $\langle \bullet \rangle$,
\begin{equation}
\langle f \left(\bm{E},t\right) \rangle \triangleq \frac{1}{2\pi} \int_{-\pi}^{\pi} f \left(\bm{E},t\right) \mathrm{d}\lambda,
\label{eq:singleavg}
\end{equation}
and apply it to both sides of \cref{eq:avg_first}, yielding
\begin{equation}
A_{i,1} = \langle F_i\left( \bm{E},t \right) \rangle.
\label{eq:avg_A1}
\end{equation}
The above equation states that, to first order, the rates of change of the mean elements are the \emph{averaged} rates of change of the osculating elements.
In the practical calculation of the $A_{i,1}$, we have to take into account their dependence from the total perturbing function $\mathcal{R} = \mathcal{R}_\Earth + \mathcal{R}'$.
Thus we make the dependence of $F_i$ from the disturbing function explicit in \cref{eq:avg_A1},
\begin{equation}
A_{i,1} = \left \langle F_i \left( \frac{\partial \mathcal{R}}{\partial \bm{E}} \left(\bm{E},t \right) , \bm{E}, t \right) \right \rangle.
\label{eq:avg_oper_A1}
\end{equation}
We can bring the averaging operator inside the parentheses since we keep the mean elements constant during the averaging operation.
Thus, the mean element rates are obtained by plugging into the osculating equations of motion the averaged total perturbing function,
\begin{equation}
A_{i,1} = F_i \left( \left \langle \frac{\partial \mathcal{R}}{\partial \bm{E}} \right \rangle, \bm{E}, t \right) = F_i \left( \frac{\partial}{\partial \bm{E}}\langle\mathcal{R}\rangle , \bm{E}, t \right) .
\label{eq:avg_pertF_A1}
\end{equation}
It can be shown that applying the averaging operator to the disturbing function is equivalent to setting $q = 2p - l$ in~\cref{eq:R,eq:R1} to eliminate the terms depending on the fast angle~\citep{Giacaglia1977}.

Additional perturbations can be superimposed by considering
\begin{align*}
\epsilon A_{i,1} &= \sum_\alpha \nu_\alpha A_{i,1\alpha}, \\
\epsilon F_{i} &= \sum_\alpha \nu_\alpha F_{i\alpha},
\end{align*}
where the index $\alpha$ varies over all the perturbations to be considered, and the $\nu_\alpha$ are the small parameters of each of the perturbations.
Each coefficient $A_{i,1\alpha}$ is calculated by averaging the corresponding perturbation,
\[
A_{i,1\alpha} = \left \langle F_{i\alpha} \left(\bm{E},t\right) \right \rangle,
\]
and coupling terms between different perturbations arise at higher orders.

Once $A_{i,1}$ is obtained through \cref{eq:avg_A1}, it is substituted in \cref{eq:mean_EoMs}, which is integrated with a suitable numerical scheme.
At each step, the osculating elements can be recovered from the mean ones by \cref{eq:osc_mean_NI}.
The short-periodic terms $\eta_{i,1}$ are computed by integrating \cref{eq:avg_first} over the mean longitude $\lambda$, while keeping the rest of the mean elements constant.

Until now, we implicitly assumed that all perturbing forces are quickly-varying (i.e., the $F_i$ are of the same order of the mean motion) through their dependence on the mean longitude $\lambda$, which is removed through the application of the single-averaging operator.
Gravitational perturbations are functions of a \emph{resonant angle} $\psi$ that is a linear combination of angles that includes $\lambda$, and which is usually quickly-varying.
However, if the orbiter is in a \emph{resonance condition}, $\psi$ changes slowly because the rates of change of $\lambda$ and of another of the angles contained in the linear combination are commensurate.
In this case, applying the single-averaging operator as in~\cref{eq:singleavg} leads to significant errors, since the long-periodic behaviors that arise from the slow variation of $\psi$ are neglected.
The issue can be solved through the application of the \emph{double-averaging} operator, which is described in the following section.

\subsubsection{Resonances}
The perturbing functions ${\distR}$ and ${\distPert}$ depend on the resonant angles $\psi_{lmpq}$ and $\psi_{lmpqp'q'}$, respectively (\cref{eq:psi_Earth,eq:psi_pert}), which are linear combinations of both fast and slow variables.
Letting $\psi$ denote either of the angles, we consider the case in which $\psi$ can be written as
\[
\psi = j{\lambda} - r{\vartheta},
\]
with $j$ and $r$ mutually prime integers.
The angle $\vartheta$ is a fast variable corresponding to either $\theta$ or $\lambda'$, depending on whether we are considering the perturbation from the tesseral harmonics of the Earth's potential or from a perturbing body, respectively.
In the latter case, we neglect the slowly-varying orbital elements of the perturbing body in the expression for $\psi$.

Let the orbiter be \emph{in resonance} if the inequality
\begin{equation}
\left| \frac{\mathrm{d}\psi}{\mathrm{d}t} \right| < \frac{2\pi}{\tau}
\label{eq:mmres_cond}
\end{equation}
is satisfied~\citep[p. 22]{Danielson1995}, where $\tau$ is the minimum period of the perturbations that must be accounted for in the averaged equations in order not to lose significant propagation accuracy.
The meaning of $\tau$ is that of an empirical ``resonance width'', which measures the maximum distance from the condition $\mathrm{d}\psi/\mathrm{d}t = 0$ for which the satellite is considered in a resonance.
The parameter $\tau$ should be chosen to be several times larger than the integration step, the mean motion, and the period of the fast angle $\tau_\vartheta$, otherwise the mean elements would include short-periodic variations.
On the other hand, $\tau$ must not be so large that long-periodic effects are inadvertently included in the short-periodic terms \citep{Morand2013}.

In the presence of a resonance, long-periodic and secular effects arise due to the commensurability of the two fast frequencies $\dot{\lambda}$ and $\dot{\vartheta}$, which pertain to the orbiter and to one of the massive bodies respectively.
By keeping $\vartheta$ fixed during the application of the single-averaging operator in \cref{eq:singleavg}, these effects are missed.
Therefore, we also integrate over $\vartheta$ by using the \emph{double-averaging operator} $\langle\langle \bullet \rangle\rangle$,
\begin{equation}
\begin{split}
\langle\langle f \left(a,h,k,P,Q,\lambda,\vartheta,t\right) \rangle\rangle \triangleq \frac{1}{4\pi^2} \int_{-\pi}^{\pi} \int_{-\pi}^{\pi} f \left(a,h,k,P,Q,\lambda,\vartheta,t\right) \mathrm{d}\lambda \mathrm{d} \vartheta + \\ + \frac{1}{2\pi^2} \sum_{(j,r) \in \mathcal{B}} \left[ \cos\psi \int_{-\pi}^{\pi} \int_{-\pi}^{\pi} f \left(a,h,k,P,Q,\lambda^*,\vartheta^*,t\right) \mathrm{d}\lambda^* \mathrm{d}\vartheta^* + \right. \\ \left. + \sin\psi \int_{-\pi}^{\pi} \int_{-\pi}^{\pi} f \left(a,h,k,P,Q,\lambda^*,\vartheta^*,t\right) \mathrm{d}\lambda^* \mathrm{d}\vartheta^* \right],
\end{split}
\label{eq:doubleavg}
\end{equation}
where $\mathcal{B}$ is the set of all $(j,r)$ for which the inequality \eqref{eq:mmres_cond} is satisfied.
Furthermore, terms depending on $\lambda$ which are responsible for the resonances have to be retained in the perturbing functions $\distEarth$  and $\distPert$.
For the case of tesseral resonances with the Earth's gravitational potential, the terms to be retained in $\distEarth$ are those satisfying the identities
\begin{align*}
    2p - l &= q - sj, \\
    m &= sr,
\end{align*}
with $s$ integer.
On the other hand, for \emph{mean-motion resonances} with the Moon, the terms to be retained in $\distPert$ obey
\[
j(l - 2p + q) = r(l' - 2p' + q').
\]
Mean-motion resonances with the Sun do not arise in practical situations due to the extremely large values of the semi-major axis that they would require.

\subsection{Averaging perturbing accelerations}
Dissipative perturbations such as atmospheric drag cannot be expressed as the gradient of a disturbing function.
In this case, the rates of change of the osculating elements can be written in Gauss form and the perturbing acceleration can be \emph{numerically averaged} over one orbital period~\citep{Uphoff1973,Ely2014}.

\subsection{Average of the short-periodic terms}
In obtaining \cref{eq:avg_A1}, we assumed that the short-periodic terms $\eta_{i,1}$ average to zero.
This is equivalent to requiring that the mean elements are ``centered'', i.e. that the short-periodic terms do not contain any long-periodic or secular offsets from the mean elements.
If one is only interested in the mean elements history this hypothesis is superfluous, and it can be overlooked.
However, in general it is necessary to impose that $\langle \eta_{i,j} \rangle = 0 $ to avoid divergence of the osculating and mean trajectories in the long term \citep{Lara2013}.

\section{Error analysis for averaged methods}
\label{sec:avg_err_analysis}
The method of averaging involves approximations needed to simplify the analytical developments, which would be intractable otherwise.
These approximations introduce errors with respect to the real trajectory that add up to the numerical error accumulated during the integration of the mean equations of motion.
While the propagation of numerical error in the integration of ordinary differential equations has been studied extensively, we are not aware of any quantitative study of the impact that the approximations involved in averaging methods have on the integration error.
In the following, we analyse the contributions to the integration error in averaged methods, and we separate them into components with distinct sources.
{Besides providing the theoretical framework for interpreting the numerical results of this study, the analysis in this section can also be used as a starting point for the improvement of existing semi-analytical methods.}

Let ${\delta}\oscE_i$ be the \emph{osculating} integration error with respect to the $i$-th equinoctial element, that is the difference between the computed osculating equinoctial element $\oscE_i$ and its true (or \emph{reference}) value.
Denoting reference values of the osculating and mean elements with a superscript $\mathrm{R}$, we have 
\begin{equation}
{\delta}\hat{E}_i = \oscE_i - \oscRefE_i. \label{eq:oscerr_def}
\end{equation}
We assume that the reference values $\oscRefE_i$ and $\bm{E}^\mathrm{R}$ satisfy \cref{eq:osc_mean_NI} exactly, while the computed values $\oscE_i$ are truncated to order $M$,
\begin{equation}
\oscE_i = E_i + \sum_{j=1}^{M} \epsilon^j {\eta_{i,j}} ( \bm{E}, t )\label{eq:osc_mean_NI_comp}.
\end{equation}
Using in \cref{eq:osc_mean_NI} the reference values,
\begin{equation}
\oscE^\mathrm{R}_i = E_i^\mathrm{R} + \sum_{j=1}^{\infty} \epsilon^j {\eta_{i,j}} \left( \bm{E}^\mathrm{R}, t \right),
\label{eq:osc_mean_NI_ref}    
\end{equation}
and subtracting the two preceding equations gives
\begin{equation}
{\delta}\hat{E}_i = \left( E_i - E^\mathrm{R}_i \right) + \left(\sum_{j=1}^{M} \epsilon^j \left( \eta_{i,j} - \eta^\mathrm{R}_{i,j} \right) - \sum_{j=M+1}^{\infty} \epsilon^j \eta^\mathrm{R}_{i,j}\right) = {\delta}E_i + {\delta}\eta_i,
\label{eq:tot_err}
\end{equation}
where ${\eta}^\mathrm{R}_{i,j} = {\eta}_{i,j}(\bm{E}^\mathrm{R},t)$.
The osculating integration error ${\delta}\hat{E}_i$ is the sum of the \emph{mean integration error} ${\delta}{E}_i = E_i - E^\mathrm{R}_i$ and of the error on the short-periodic terms ${\delta}{\eta}_i$.
The mean integration error can be further decomposed by considering
\begin{equation}
{\delta}{E_i} = {\delta}{E}_{i, \mathrm{dyn}} + {\delta}{E}_{i, \mathrm{mod}} + {\delta}{E}_{i, \mathrm{num}},
\label{eq:avg_err_breakdown}
\end{equation}
where the first term is the \emph{dynamical error}, the second is the \emph{model truncation error}, and the last is the \emph{numerical integration error}.

\subsection{Dynamical error \texorpdfstring{$\delta{E}_{i,\mathrm{dyn}}$}{dEdyn}}
\label{sec:dyn_err}
For a first-order theory, the second term in \cref{eq:mean_EoMs} is truncated at $j=1$ yielding
\begin{equation}
\dt{E_i} = n(a) \delta_{i,6} + \epsilon A_{i,1}.
\label{eq:mean_EoM_trunc}
\end{equation}
The mean rate $A_{i,1}$ is obtained by \cref{eq:avg_oper_A1}, in which the mean elements are kept constant during the averaging operation.
We define the reference mean rates $A^\mathrm{R}_{i,1}$ as those that would be obtained by computing exactly the definite integrals in \cref{eq:singleavg,eq:doubleavg}, i.e. by taking into account the time dependence of the mean orbital elements over the period of integration.
Plugging into \cref{eq:mean_EoMs} the exact mean rates yields
\begin{equation}
\frac{\mathrm{d}{{E}^\mathrm{R}_i}}{\mathrm{d}t} = n\left(a \right) \delta_{i,6} + \sum_{j=1}^{\infty} \epsilon^j A^\mathrm{R}_{i,j},
\label{eq:mean_EoMs_exact}
\end{equation}
which, when subtracted from~\cref{eq:mean_EoM_trunc}, gives the error on the total mean rates of change
\begin{equation}
\delta{\dot{E}_{i,\text{dyn}}} = \epsilon \left( A_{i,1} - A^\mathrm{R}_{i,1} \right) - \sum_{j=2}^{\infty} \epsilon^j A^\mathrm{R}_{i,j} = \epsilon {\delta} A_{i,1} + \mathcal{O}\left(\epsilon^2 \right).
\label{eq:dyn_err}
\end{equation}
This error is the sum of a term of order $\epsilon$, due to the inexact result from the application of the averaging operator, and higher-order terms that are neglected.
At each integration step, we write the \emph{dynamical error} can be quantified by
\begin{equation*}
\delta{E}_{i,\text{dyn}} = \delta{\dot{E}_{i,\text{dyn}}} \Delta t.
\end{equation*}

From the standpoint of numerical integration, $\delta{E}_{i,\text{dyn}}$ propagates during the integration in a way similar to the local truncation error.
However, the dynamical error cannot be reduced by choosing a smaller time step.
Situations giving rise to a large $\delta{E}_{i,\text{dyn}}$ take place if the mean elements change significantly during one orbital period, or if higher orders are non-negligible in the equations of averaging~(\cref{eq:averaging}).
This would be also the case if the single-averaging operator was applied in the presence of a resonance, without taking into account that the slow rate of change of the resonant angle $\psi$ generates long-periodic or secular trends.

In \cref{eq:dyn_err}, we have only considered first-order mean rates of a single perturbation.
The development carried out here can be extended to higher orders and applied to several perturbations, which will give rise to further terms in the definition of the dynamical error.

\subsection{Model truncation error \texorpdfstring{${\delta}E_{i, \mathrm{mod}}$}{dEmod}}
To perform the analytical averaging of conservative perturbations, the expressions of the perturbing functions in \cref{eq:R,eq:R1} and the eccentricity functions $K_{lpq}$, $L_{lpq}$ and $G_{lp'q'}$ have to be truncated at a given order $l_\mathrm{M}$.
Also, the only terms retained in the sums on $q$ and $q'$ are those that give rise to long-periodic effects, which we denote with the finite sets $\mathcal{Q}$ and $\mathcal{Q'}$.
In this way, only the lowest frequencies governing the motion are taken into account, reducing computational complexity and cost.
By defining $\distEarth^\mathrm{T}$ and $\mathcal{R'}^\mathrm{T}$ as the truncated perturbing functions,
\begin{align}
\distEarth^\mathrm{T} &= \sum_{l=2}^{l_\mathrm{M}} \sum_{m=0}^{l} \sum_{p=0}^{l} \sum_{q \in \mathcal{
    Q}} \mathcal{R}_{lmpq} \label{eq:Rtrunc} \\
\mathcal{R'}^\mathrm{T} &= \sum_{l=2}^{l_\mathrm{M}} \sum_{m=0}^{l} \sum_{p=0}^{l} \sum_{p'=0}^{l} \sum_{q \in \mathcal{Q}} \sum_{q \in \mathcal{Q'}} \distPert_{lmpqp'q'} \label{eq:R1trunc},
\end{align}
we find the truncation error on the perturbing function $\delta\mathcal{R}$ as
\[
\delta\mathcal{R} = ( \distEarth^\mathrm{T} - \distEarth ) + ( \mathcal{R'}^\mathrm{T} - \distPert ).
\]
Since the mean rate ${A}_{i,1}$ is obtained by averaging $\mathcal{R}^\mathrm{T}$, the error $\delta\mathcal{R}$ will give rise to a \emph{model truncation error} $\delta{E_{i, \text{mod}}}$ on the mean orbital elements, which will propagate during the integration of the equations of motion, analogously to $\delta{E_{i,\text{dyn}}}$.
The model truncation error is caused by neglecting higher-order terms in the perturbing functions.
In fact, one may build high-order theories (in the mean rates $A_{i,j}$) while keeping only a limited number of the most relevant terms in the perturbing functions.

Note that the gravitational perturbing function of the geoid is always truncated even for non-averaged methods, since a closed form does not exist.
However, the Sun and the Moon are usually considered as point masses, an approximation that holds well for Earth satellite orbits.
Thus, their perturbing function can be expressed in a closed form that can be evaluated efficiently by non-averaged methods~\citep[section 8.4]{Battin1999}.

\subsection{Numerical error \texorpdfstring{$\delta{E}_{i,\mathrm{num}}$}{dEnum}}
The mean elements at each step of the numerical integration of \cref{eq:mean_EoMs} are affected by truncation and round-off errors.
Dropping the subscript $i$ for ease of notation, we denote them with ${\delta}{E}_\mathrm{num, T}$ and ${\delta}{E}_\mathrm{num, R}$ respectively,
\[
{\delta}{E}_\mathrm{num} = {\delta}{E}_\mathrm{num, T} + {\delta}{E}_\mathrm{num, R}.
\]
If the numerical scheme is of $p$-th order, the local truncation error accumulated with a step size ${\Delta}t$  is proportional to the $(p+1)$-th derivative of the mean element,
\[
{\delta}E_{\mathrm{num, T}} \sim \frac{\mathrm{d}^{p+1} E}{\mathrm{d} t^{p+1}}  {\Delta}t^{p+1}.
\]

The round-off error ${\delta}{E}_\mathrm{num, R}$ grows with the number of floating-point operations performed during the integration. Thus, it is proportional to some power of the total number of integration steps.
The accumulated round-off error can be estimated through statistical laws~\citep{Brouwer1937,Milani1987}.
Note that in non-averaged methods the osculating integration error (\cref{eq:tot_err}) is only due to the numerical error.

\subsection{Error on the short-periodic terms \texorpdfstring{$\delta{\eta}_i$}{deltaeta}}
\label{sec:error_sp}
Following \cref{eq:tot_err}, the error on the short-periodic terms can be written as
\begin{equation}
\delta{\eta_{i}} = \sum_{j=1}^{M} \epsilon^j ( \eta_{i,j} - \eta^\mathrm{R}_{i,j} ) - \sum_{j=M+1}^{\infty} \epsilon^j \eta^\mathrm{R}_{i,j}.
\label{eq:err_sp}
\end{equation}
We expand $( \eta_{i,j} - \eta^\mathrm{R}_{i,j} )$ as
\[
\eta_{i,j}(\bm{E},t ) - {\eta}_{i,j}(\bm{E}^\mathrm{R},t ) \approx  \frac{\partial \eta_{i,j}}{\partial E_i}(\bm{E}^\mathrm{R},t) \delta{E_i},
\]
where $\delta{E_i}=E_i-E_i^\mathrm{R}$.
Substituting in \cref{eq:err_sp} we obtain
\begin{equation}
\delta{\eta}_i \approx \sum_{j=1}^{M} \epsilon^j \frac{\partial {\eta}_{i,j}}{\partial {E}_i} \delta{E_i} - \sum_{j=M+1}^{\infty} \epsilon^j \eta^\mathrm{R}_{i,j} = \delta{\eta}_{i,\mathrm{M}} + \delta{\eta}_{i,\mathrm{HO}}.
\label{eq:err_sp_break}
\end{equation}
The error on the short-periodic terms is given by two contributions.
The first, $\delta{\eta}_{i,\mathrm{M}}$, is due to the mean integration error $\delta {E}_i$ as defined above.
In fact, the short-periodic terms are computed from values of the mean elements that are affected by $\delta {E}_i$, generating an error on the short-periodic terms up to the $M$-th order.
The second contribution, $\delta{\eta}_{i,\mathrm{HO}}$, is due to the truncation to the $M$-th order that is performed in \cref{eq:osc_mean_NI_comp}.
The first term of $\delta{\eta}_i$ is proportional to $\epsilon \delta{E}_i$, and will be negligible compared to the other sources of error.

The impact of the error on the short-periodic terms is particularly critical for the initial conditions of a propagation, which are often assigned in osculating elements.
The \emph{mean} initial conditions for a semi-analytical propagation must be derived by subtracting the short-periodic terms, which are affected by the error $\delta{\eta}_i$.
The error on the mean initial conditions will lead to a divergence of the computed trajectory from the reference which might become critical in some cases, for example when accuracy requirements are stringent or the dynamics is chaotic. 

\subsection{Error budget}
It is desirable that the dynamical error ${\delta}E_{i,\mathrm{dyn}}$ dominates over the remaining terms. 
This is because both the model truncation error ${\delta}{E_{i,\mathrm{mod}}}$ and the numerical error ${\delta}{E_{i,\mathrm{num}}}$ can be driven down, within certain limits. Assuming that the analytical expressions for the expansions are available, the error ${\delta}{E_{i,\mathrm{mod}}}$ for the lunar and solar perturbing functions can be reduced by choosing higher truncation indices in \cref{eq:Rtrunc,eq:R1trunc}.
Moreover, it is possible to reduce ${\delta}{E_{i,\mathrm{num}}}$ by using an efficient numerical scheme, and by applying an appropriate value for the tolerance or the step-size of the solver.
As we will show in the numerical tests, it is generally possible to decrease ${\delta}{E_{i,\mathrm{num}}}$ so that it is negligible with respect to the other error contributions.
Situations in which the numerical error dominates might only arise in extremely long propagations, for which round-off error could play a significant role.

\section{Orbit propagation software}
\label{sec:software}
In the following, we outline the characteristics of the software used to perform the numerical tests.
To ensure the validity of the analysis performed on the results from the numerical tests, it is necessary to carefully \emph{align} the codes, i.e., to implement the same physical model in order to exclude any difference in numerical results due to different modeling of the physical phenomena.

\subsection{\STELA}
The {\STELA} (Semi-analytical Tool for End of Life Analysis) orbit propagator has been developed by the French space agency CNES to assess compliance with spacecraft re-entry requirements imposed by the French Space Act~\citep{LeFevre2014}, and is similar in function to the DSST.
The software is publicly available as a Java executable,\citeurl{https://logiciels.cnes.fr/content/stela?language=en}{May 31\textsuperscript{st}, 2018} and in this work we use version 3.2.

{\STELA} integrates single-averaged equations of motion for the equinoctial element set $\bm{E} = (a,h,k,P,Q,\lambda)$ with a Runge-Kutta solver with fixed step size $\Delta{t}$.
The order of integration can be chosen between \num{4} and \num{6}; we use the latter value for all of our simulations.
We assign the initial conditions for the propagations in osculating orbital elements.
{\STELA} automatically removes the short-periodic terms from the initial osculating elements in order to obtain the mean initial conditions for the propagation.

\subsubsection{Physical model}
\label{sec:STELA_mathmod}
{\STELA} allows the user to choose which perturbations to include among several options.
For the conservative perturbations, we take into account the geopotential corresponding to the \textsc{GRIM5-S1} model~\citep{Biancale2000}, and a simplified version of the analytical solar and lunar ephemerides by \citet{Meeus1998}.
The geopotential harmonics can be computed up to any degree and order through a recurrence formulation.

{\STELA} truncates the expansion~\eqref{eq:Rtrunc} of the Earth's non-spherical perturbing function at the second degree in the presence of tesseral resonances~\citep{Morand2013}.
\Cref{eq:R1trunc}, that is the expansion of the solar and lunar perturbing functions, is truncated at an order $l_\mathrm{M}$ that can be varied between 2 and 8 by the user.
The short-periodic terms (\cref{eq:osc_mean_NI_comp}) are computed at first order for lunisolar perturbations and at second order for $J_2$.
Resonant tesseral harmonics are retained in the perturbing function according to their period, that is $2\pi / \dot{\psi}_{lmpq}$ with $\psi_{lmpq}$ defined according to \cref{eq:psi_Earth}.
If the period of a tesseral harmonic is greater than a customizable multiple $N_\mathrm{tess}$ of the integration step, it is retained in the averaged perturbing function~\citep{Morand2013}; that is, the tesseral harmonics retained according to condition \eqref{eq:mmres_cond} have periods greater than $\tau_\mathrm{tess} = N_\mathrm{tess} \Delta{t}$.
While {\STELA} considers precessional and nutational movements of the rotational axis of the Earth by integrating the equations of motion in the Celestial Intermediate Reference Frame (CIRF)~\citep{STELA_Man2016}, we disabled these effects in our tests as they are not relevant for the evaluation of the numerical performance.

\STELA{} can also consider drag arising from an atmosphere co-rotating with the Earth as an additional perturbation.
As to obtain the corresponding contribution to the rate of change of the mean elements (\cref{eq:mean_EoMs}), first the osculating elements are recovered at a prescribed number of points $M_\mathrm{quad}$ along the orbit.
The drag acceleration is computed from the osculating elements at each of these points, and its average over one period is performed through a numerical quadrature.
This algorithm is executed every $N_\mathrm{drag}$ integration steps; both $N_\mathrm{drag}$ and $M_\mathrm{quad}$ can be chosen by the user depending on the required balance between speed and accuracy.
The atmospheric density is computed using either the US 1976 Standard Atmosphere (US76), the Jacchia 1977, or the NRLMSISE-00 models~\citep{US76,Jacchia1977,Picone2002}.
To better align the codes, we assume the solar flux at the \SI{10.7}{\centi\metre} wavelength and the planetary geomagnetic amplitude $A_\mathrm{p}$ to be equal to 140 SFU and 15, respectively.
We neglect the Earth oblateness in the computation of the geodetic height by setting the ellipticity to zero.
\STELA{} only computes the drag acceleration when the altitude is lower than an assigned value of \SI{1000}{\kilo\metre}.


\STELA{} also includes perturbations stemming from Earth solid tides and solar radiation pressure; however, we turn them off for this study.


\subsection{\Thalassa}
\Thalassa{} is an orbit propagation code that numerically integrates the non-averaged equations of motion of the perturbed two-body problem written for different formulations{; the code was specifically developed for the present study.}
{\Thalassa} includes the classical Cowell's method \citep[][p. 447]{Battin1999}, in which the Cartesian coordinates of the position and velocity relative to the primary body of attraction are employed as state variables, and three regularized formulations.
The Kustaanheimo Stiefel (KS) regularization has been implemented following \citet[][pp. 33-35]{Stiefel1971}. The independent variable $s$ is defined by the time transformation (known also as the Sundman transformation)
\begin{equation}
\frac{d t}{d s}=\frac{r}{\beta},
\label{eq:Sund}
\end{equation}
where $\beta=1$, $t$ is the physical time, and $r$ is the orbital distance.
For negative values of the total energy $h$, the two-body problem is transformed into four uncoupled harmonic oscillators of equal frequency $\sqrt{-h/2}$. 
The state vector is composed by the four KS parameters along with their derivatives with respect to $s$, the total energy, and time.
In our implementation, we also consider a linear time element instead of time as a dependent variable, because a better performance can be reached for more eccentric orbits.
By analytically integrating \cref{eq:Sund} when the motion is unperturbed a time element can be introduced as either a constant quantity or a linear function of $s$ \citep[see][]{Bau2014}. 

We also implemented in {\Thalassa} two formulations based on regular elements (see the Introduction).
The first is a VOP method related to the KS scheme which was developed by \citet{Broucke1966}.
Eight elements are obtained from the solution of the KS harmonic oscillators for $h<0$, and the total energy and a linear time element complete the state vector.
These elements are non-singular for any inclination and eccentricity smaller than 1 and are well-defined at collision.
We refer to this method as SS because we implemented the same equations derived in \citet[][section 19]{Stiefel1971}.
{The second one, named EDromo, has been recently proposed by \citet{Bau2014a,Bau2015}.
This formulation was inspired by the ideal elements first proposed by \citet{Deprit1975} and later on by \citet{Pelaez2007}, where their connection to the regularization devised by \citet{Burdet1969} and
\citet{Ferrandiz1988} becomes clear. For a comprehensive overview of the ideal elements we refer to the
Introduction of \citet{Bau2015} and to \citet{Lara2017a}. The basic idea is to decompose the motion into
the rotation of the radial unit vector and the displacement along the radial direction. In EDromo
four elements are the Euler parameters that define the orientation of an intermediate reference frame
in space.}
This frame is fixed when there are no perturbations and has one axis pointing towards the angular momentum vector.
Other three elements determine the shape of the ellipse and together with the fictitious time allow us to locate the particle along the orbit.
Either a linear or a constant time element can be used.
EDromo elements are non-singular like the SS ones, but they do not work at collision.
For both SS and EDromo the independent variable is related to time by \cref{eq:Sund} with $\beta=\sqrt{-h/2}$. \Cref{tab:forms} summarizes some relevant features of the formulations contained in {\Thalassa}.

\begin{table}[t]
\centering
\caption{For the formulations implemented in \Thalassa{} we specify if the spatial variables consist in orbital elements, i.e. constant quantities along the Keplerian solution, or not, if a time element is employed, and the dimension of the state vector.}\vspace{0.2in}
\begin{tabular}{ l c c c }
    \toprule
    Formulation & Variables & Time element & Dimension\\
    \midrule
    Cowell & coordinates & no   & 6 \\
    KS     & coordinates & yes  & 10 \\
    SS     & elements    & yes  & 10 \\
    EDromo & elements    & yes  & 8 \\
    \bottomrule
\end{tabular}
\label{tab:forms}
\end{table}

The equations of motion are integrated with the {\LSODAR} numerical solver\citeurl{http://www.netlib.org/odepack/}{October 15\textsuperscript{th}, 2017} \citep{Radhakrishnan1993} in double precision.
{\LSODAR} chooses the step size and order along the integration according to relative and absolute tolerances on the local truncation error, which are set equal to a single value assigned by the user.
In all the test cases, we use very strict values (\num{e-18} or smaller) of the {\LSODAR} local truncation error tolerances to generate reference trajectories in quadruple precision.

\subsubsection{Physical model}
\Thalassa{} has been aligned to \STELA{} by implementing the same set of perturbations, and by using the same sources for the physical constants, ephemerides, and atmospheric models.
The perturbations included in {\texttt{THA\-LASSA}} are the same as those considered in {\STELA}, and they are taken from the same models.\footnote{The subroutine for the computation of the ephemerides was kindly provided by Florent Deleflie.}
All the constants involved in the computations for the geopotential and lunisolar perturbations are identical, as to exclude any difference in the computed trajectories due to different perturbation models.
The code for the US76 model is ported from the one used in \STELA{}, while the NRLMSISE-00 model is implemented through the official subroutines that are publicly available.\citeurl{https://ccmc.gsfc.nasa.gov/pub/modelweb/atmospheric/msis/nrlmsise00/}{October 15\textsuperscript{th}, 2017}
The value of the air density is replicated to double precision machine zero with the US76 model.
We could not assure a similar level of alignment in the computation of density with the Jacchia 77\citeurl{http://www.dem.inpe.br/~val/atmod/default.html}{October 15\textsuperscript{th}, 2017} and NRLMSISE-00 models.
This is because the models only prescribe the fundamental equations, constants and analytical procedures to compute the air density, but several choices regarding their implementation are left to the user.
Since the source code of \STELA{} is not available to the public, it is not possible to replicate these choices on a line-by-line basis, preventing reproduction of the results at machine precision.

A particularly important section of the {\Thalassa} code involves the calculation of the non-spherical gravity perturbation.
From a preliminary test on a LEO orbit, we discovered that the evaluation of this perturbation can absorb more than 50\% of the computational time for a moderately complex (e.g., $5 \times 5$) model.
In order to reduce this burden, we implement the algorithm for the calculation of the geopotential harmonics by \citet{Pines1973}.
The algorithm is quite efficient in that only two trigonometric function evaluations are required for each evaluation of the perturbing function.
Besides, no singularities at the poles are present, in contrast to the explicit evaluation of the Legendre associated functions in the classical perturbing function expansion.


{Lunisolar gravitational perturbations are implemented in \Thalassa{} according to two different mathematical approaches.}
The first approach consists of the classical third-body perturbing acceleration $\bm{P}'$ expressed as the derivative of the perturbing function $\mathcal{R}'$ written in Cartesian coordinates~\citep[section 8.4]{Battin1999},
\begin{align}
    \mathcal{R}' &= \mu' \left( \frac{1}{d} - \frac{\bm{r} \cdot \bm{r}'}{rr'} \right) \label{eq:R1Cart}, \\
    \bm{P}' &= \frac{\partial \mathcal{R'}}{\partial \bm{r}} = -\mu' \left( \frac{\bm{d}}{d^3} + \frac{\bm{r}'}{\left(r'\right)^3} \right), \label{eq:P1Cart}
\end{align}
where $\bm{r}'$ is the position vector of the perturbing body with respect to the Earth, and $\bm{d} = \bm{r} - \bm{r}'$.
In order to isolate the effect of the model truncation error ${\delta}E_{i, \mathrm{mod}}$ from that of the dynamical error $\delta{E}_{i,\mathrm{dyn}}$, in our second approach we consider the perturbing acceleration derived from the \emph{truncated} perturbing function $\mathcal{R'}^\mathrm{T}$.
Expanding $\mathcal{R}'$ in Legendre polynomials gives
\begin{equation}
    \mathcal{R'} = \frac{\mu}{r'} \left[ 1 + \sum_{l=2}^{\infty} \left( \frac{r}{r'} \right)^l P_l \left( \frac{\bm{r} \cdot \bm{r}'}{rr'} \right) \right], \label{eq:R1Legendre}
\end{equation}
where $P_l \left( \nu \right)$ is the Legendre polynomial of order $l$,
\begin{equation}
    P_l \left( \nu \right) \triangleq \sum_{s=0}^{\left \lfloor \frac{1}{2} l \right \rfloor} \frac{\left( -1 \right)^s \left( 2l - 2s \right)!}{2^l s! \left( l - s \right)! \left( l - 2s \right)!} \nu^{l - 2s} = \sum_{s=0}^{\left \lfloor \frac{1}{2} l \right \rfloor} G(l,s) \nu^{l - 2s}.
\end{equation}
This expression for $\mathcal{R'}$ is formally equivalent to that in \cref{eq:R1}, and in fact it is the first step in the derivation of $\mathcal{R}'$ as a function of the classical orbital elements~\citep[section 6.3]{Murray1999}.
By making $P_l$ explicit in \cref{eq:R1Legendre} and manipulating the resulting expression we rewrite $\mathcal{R'}$ as
\begin{equation}
    \mathcal{R'} = \frac{\mu}{r'} \left[ 1 + \sum_{l=2}^{\infty} \sum_{s=0}^{\left \lfloor \frac{1}{2} l \right \rfloor} \frac{G(l,s)}{(r')^l} r^{2s} \left( \bm{r} \cdot \bm{\hat{r}'} \right)^{l-2s} \right], \label{eq:R1Gls}
\end{equation}
where $\bm{\hat{r}'} = \bm{r}'/r'$ is a unit vector. By taking the gradient of \cref{eq:R1Gls}, truncating the expansion at order $l_\mathrm{M}$, and rearranging the terms, we obtain a compact expression for ${\bm{P}'}^\mathrm{T}$, the lunisolar perturbing acceleration resulting from the truncated perturbing function $\mathcal{R'}^\mathrm{T}$,
\begin{equation}
    {\bm{P}'}^\mathrm{T} = \frac{\partial \mathcal{R'}^\mathrm{T}}{\partial \bm{r}} = \frac{\mu'}{r'} \sum_{l=2}^{l_\mathrm{M}} \sum_{s=0}^{\left \lfloor \frac{1}{2} l \right \rfloor} \frac{G(l,s)}{(r')^l} r^{2s} \left( \bm{r} \cdot \bm{\hat{r}'} \right)^{l-2s} \left[ 2s \frac{\bm{r}}{r^2} + \frac{\left( l - 2s\right)}{ \left( \bm{r} \cdot \bm{\hat{r}'} \right)} \bm{\hat{r}'}  \right]. \label{eq:P1Trunc}
\end{equation}
Using a low-order expansion for ${\bm{P}'}^\mathrm{T}$, rather than $\bm{P}'$, can increase numerical accuracy when the orbit is very close to the primary body~\citep[p. 388]{Battin1999}.
However, by expressing lunisolar perturbations through ${\bm{P}'}^\mathrm{T}$ we aim to introduce an ``artificial'' model truncation error.
For equal $l_\mathrm{M}$, \Thalassa{} propagations performed with ${\bm{P}'}^\mathrm{T}$ will be affected by the same model truncation error as \STELA{}, thus allowing to isolate its effect on the total integration error separately from the other contributions.
The numerical test cases in \cref{sec:GTO_test,sec:HEO_test} highlight these considerations.
In the LEO test case, we always consider the full expression $\bm{P}'$.

\subsubsection{{\Thalassa} implementation and comparison with \STELA{}}
\label{sec:thalassa_impl}
Any mathematical procedure implemented in a computer may exhibit very different computational cost according to implementation\hspace{0pt}-dependent features such as the memory access method, the number of disk I/O operations, and source code language, factoring and optimization.
It is important to carefully choose these features in order to attain the maximum performance.
In the following, we provide a broad overview of the implementation of the {\Thalassa} code.

The main {\Thalassa} executable is implemented in Fortran 2008, and is available through a public Gitlab repository.\footnote{\textsc{URL:} \url{https://gitlab.com/souvlaki/thalassa}, last visited May 31\textsuperscript{st}, 2018.}
The Fortran code is written for sequential execution, but the repository also includes Python scripts that execute propagations in parallel over several cores for large scale simulations.
We use the \Thalassa{} version 1.2 for the tests.
For the purposes of comparing the computational time with respect to \STELA{} we emphasize that, depending on multiple factors such as the machine architecture, Java Development Kit and Fortran compiler versions, programs written in Fortran may exhibit lower computational times than their equivalents in Java.
Benchmarks for scientific computing~\citep{Bull2001} and applications to astrodynamics~\citep{Eichhorn2017} show that Java codes are usually slower than Fortran ones, although the computational time strongly depends on the particular benchmark and on the characteristics of the implementation.
The factor by which Java algorithms are slowed down with respect to their Fortran versions can be conservatively estimated as 5 for our applications.
Since \STELA{} is heavily optimized for performance and its internal algorithms have not been released publicly, in the rest of this article we will report the actual CPU times required to run the different codes, without assigning any ``penalty'' to \Thalassa{}.

The \Thalassa{} code is compiled by using \texttt{gfortran 6.3.0} with the optimization flag \texttt{-O}; the tests are performed on an iMac Pro equipped with 18 \SI{2.3}{\giga\hertz} Intel Xeon W cores with up to \SI{4.3}{\giga\hertz} Turbo Boost.


\section{LEO numerical test case}
\label{sec:LEO_test}
\begin{table}[t]
\centering
\caption{Initial modified Julian date and osculating orbital elements for the simulated Tintin A spacecraft.
The last row refers to the osculating argument of longitude $\hat{u} = \hat{\omega} + \hat{M}$.\label{tab:ICs_LEO}}
\[
    \begin{array}{cS[tight-spacing=true,table-format=5.6e0]s}
    \toprule
    \text{MJD}            & 58171.738177  &             \\
    \hat{a}               & 6892.14       & \kilo\metre \\
    \hat{e}               & 0             &             \\
    \hat{i}               & 97.46         & \degree     \\
    \hat{\Omega}          & 281           & \degree     \\
    \hat{u}               & 0.0           & \degree     \\
    \bottomrule
    \end{array}
\]
\end{table}
We consider the propagation of a LEO orbit corresponding to that of the \emph{Tintin A} spacecraft.\citeurl{http://space.skyrocket.de/doc_sdat/microsat-2.htm}{July 25\textsuperscript{th}, 2018}
The initial osculating orbital elements in \cref{tab:ICs_LEO} correspond to its parking, Sun-synchronous orbit about \num{30} minutes after launch on February 22\textsuperscript{nd}, 2018.

We assume a constant drag coefficient $C_\mathrm{D} = 2.2$, a spacecraft mass of \SI{400}{\kilo\gram} and a cross-sectional area of \SI{0.7}{\square\metre}.
The physical model includes a $5 \times 5$ geopotential, lunisolar perturbations and atmospheric drag.
The solar flux and the geomagnetic planetary index and amplitude are kept constant ($K_p = \num{3.0}$, $A_p = \num{15}$, $F_{10.7} = \SI{140}{\SFU}$).
\begin{figure}[t]
\centering
\includegraphics[width=0.8\columnwidth]{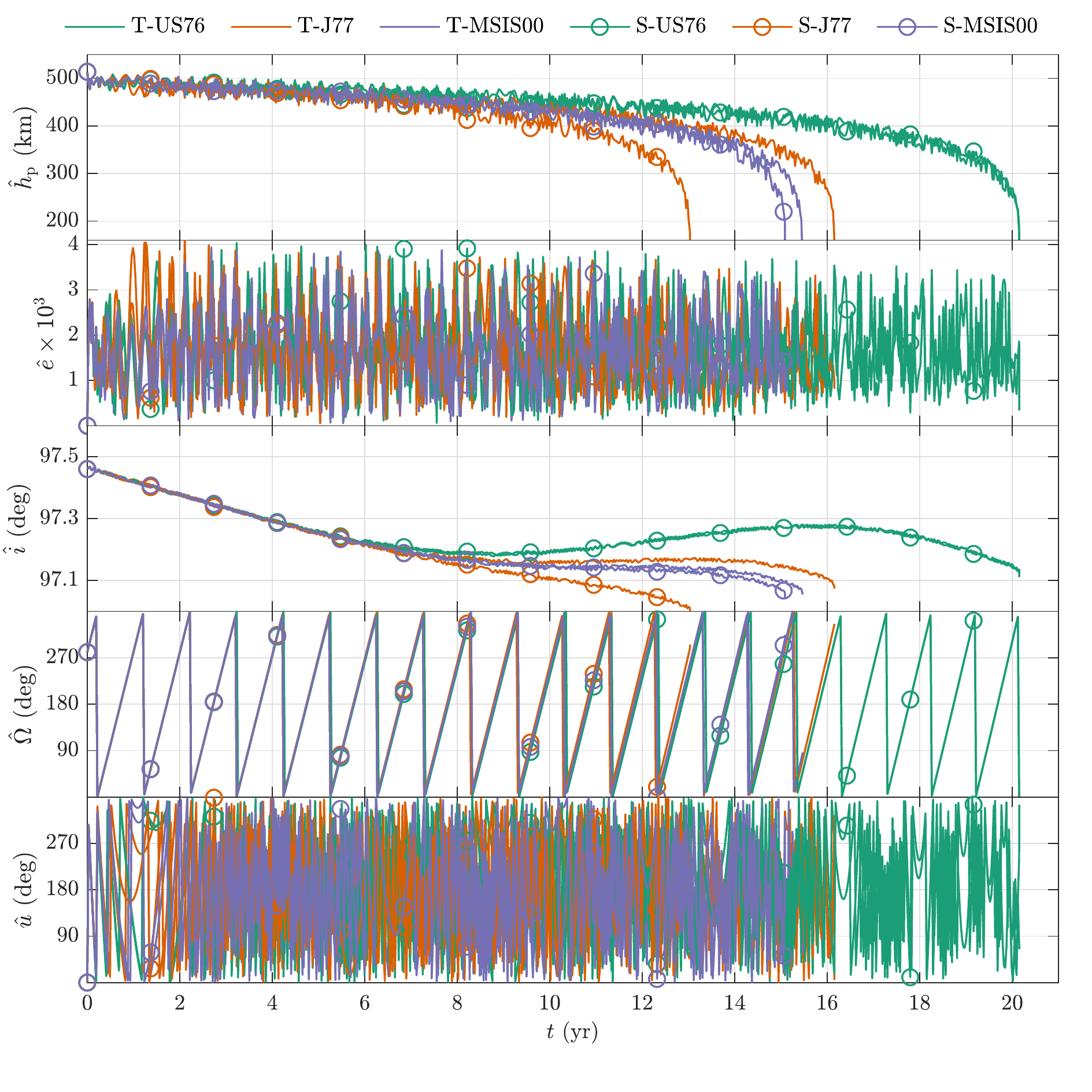}
\caption{Osculating perigee altitude $\hat{h}_\mathrm{p}$ and remaining orbital elements as a function of time for the propagation of the \emph{Tintin A} initial conditions in \cref{tab:ICs_LEO} until re-entry due to atmospheric drag (the mean argument of latitude is $\hat{u} = \hat{M} + \hat{\omega}$).
Green, orange, and purple curves are obtained with the US76, Jacchia 77, and NRLMSISE-00 atmospheric models, respectively.
The trajectories are computed with both \Thalassa{} and \STELA{}; the latter are shown with circular markers.
\label{fig:LEO_COE}}
\end{figure}

\begin{table}[t]
\centering
\caption{Values of the parameters used in the \STELA{} propagation for the LEO test case. The meaning of the parameters is explained in \cref{sec:STELA_mathmod}. The order $l_\mathrm{M}$ refers to the expansions of the lunar and solar perturbing functions. \label{tab:STELA_LEO_params}}
\[
\begin{array}{c*5{S[tight-spacing=true]}}
\toprule
{\Delta {t}} & {l_\mathrm{M}} & {N_\mathrm{tess}} & N_\mathrm{drag} & {M_\mathrm{quad}} \\
\midrule
\SI{24}{\hour} & 3 & 20 & 1 & 33 \\
\bottomrule
\end{array}
\]
\end{table}

\subsection{LEO orbital dynamics}
\Cref{fig:LEO_COE} shows the history of the osculating perigee altitude and of the remaining osculating orbital elements for the propagation of the initial conditions in \cref{tab:ICs_LEO} until a re-entry at the height of \SI{120}{\kilo\metre} is detected.
We display the trajectories obtained with both \Thalassa{} and \STELA{} for all the atmospheric models.
The parameters affecting the numerical propagation in \STELA{} are chosen through a trial-and-error calibration procedure.
Their nominal values, shown in \cref{tab:STELA_LEO_params}, are chosen as those for which the final re-entry date converges within an acceptable computational time.
The \Thalassa{} trajectories are propagated by integrating the KS equations with a solver tolerance of \num{e-14} and no time element.

Examination of \cref{fig:LEO_COE} shows that the choice of atmospheric model heavily impacts the lifetime estimate of $\num{16.6} \pm \num{3.6}$ years.
In fact, the modelling of atmospheric drag constitutes the largest physical source of uncertainty for LEO orbits.
Curves for \STELA{} and \Thalassa{} almost overlap when using the US76 atmospheric model, as the values of atmospheric density are consistent in both codes within a few units of double precision machine zero.
With this model, existing discrepancies are entirely to be attributed to the dynamical and short-periodic errors in \STELA{}, since the numerical error in the \Thalassa{} solution is substantially mitigated due to the very small tolerance value.

\subsection{Performance of the semi-analytical method}
\begin{figure}
\centering
\includegraphics[scale=0.5,clip,trim={0in 0.2in 0in 0.6in}]{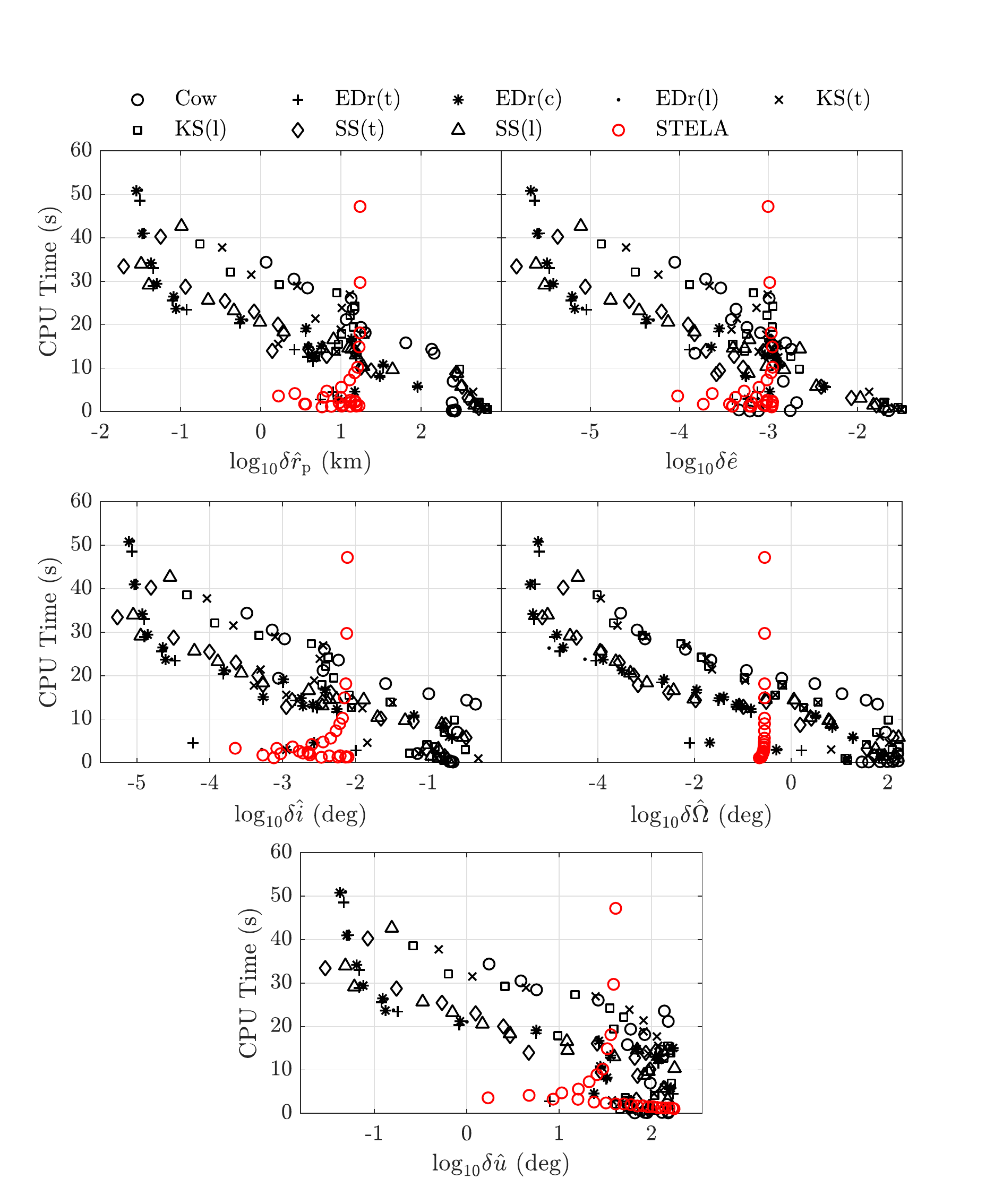}
\caption{Measured CPU time as a function of the errors on the osculating orbital elements for the LEO test case with the US76 atmospheric model, with respect to a reference solution in quadruple precision. Error values are obtained by changing the solver tolerance (for \Thalassa{}, black markers) or the solver step size (for \STELA{}, red markers), and are all measured after 18 years of propagation of the initial conditions in \cref{tab:ICs_LEO}.
The labels of the \Thalassa{} data series refer to the Cowell formulation (``Cow''), EDromo (``EDr''), Kustaanheimo Stiefel regularization (``KS''), and Stiefel-Scheifele set of elements (``SS'').
In the regularized formulations, the physical time is computed by either integrating the time transformation (``(t)''), or from a linear and a constant time element (``(c)'' and ``(l)'', respectively).\label{fig:LEO_batch_COE}}
\end{figure}
\label{sec:LEO_SA_perf}
\Cref{fig:LEO_batch_COE} shows the CPU time as a function of the errors in osculating orbital elements with respect to a reference trajectory computed by running \Thalassa{} in quadruple precision and with a solver tolerance of \num{e-16}.
The error values refer to trajectories obtained by varying the \Thalassa{} solver tolerance between \num{e-3} and \num{e-13}, and the \STELA{} integration ${\Delta}t$ between \SI{0.01}{\day} and \SI{3}{\day}.
The CPU times are averaged over 3 runs of each propagation.
We evaluate the errors at a common epoch of \num{18} years from the start rather than at the re-entry epoch, since the latter changes according to different propagations.
In both codes, we consider the US76 atmospheric model to eliminate any sources of error different from those described in \cref{sec:avg_err_analysis}.

\STELA{} attains \num{10} \si{\kilo\meter} accuracy in the radius of perigee with CPU times between \num{5} and \num{50} seconds, depending on the time step.
Excluding very small time step values, the average \STELA{} computational time is in the order of \num{5} seconds.
{\Thalassa} requires larger CPU times for the same accuracy, between \num{10} to \num{30} seconds according to the chosen formulation.
In this respect, regularized element methods such as EDromo and the one due to Stiefel and Scheifele achieve the smallest computational times, which is twice that obtained with \STELA{}.
Note that previous works estimate non-averaged methods to be a hundred times slower than semi-analytical ones~\citep{Lara2012}.

However, the maximum accuracy attainable by \STELA{} is limited by the approximations intrinsic to the averaging process.
While the expansions of the gravitational perturbing functions converge very quickly due to the small eccentricity, the mean integration error ${\delta}{E}_i$ (\cref{eq:avg_err_breakdown}) still impacts the total integration error considerably.
It can be shown that the latter is one order of magnitude larger than the short-periodic terms in the eccentricity and angular variables.
Further \STELA{} propagations performed with $l_\mathrm{M} > 3$ do not result in significant improvements, suggesting that the lunisolar model truncation error is negligible, as it can be expected from the small value of the semi-major axis. 

As mentioned in \cref{sec:avg_err_analysis}, the dynamical error, the model truncation error, and the error on the short-periodic terms  are entirely independent from the numerical integration process and it is not possible to mitigate them by choosing a smaller step size.
According to \cref{eq:avg_err_breakdown}, the time step ${\Delta}t$ should only be small enough to ensure that $\left|{\delta}E_\mathrm{num}\right| \ll |{\delta}E_\mathrm{dyn}| + |{\delta}E_\mathrm{mod}|$.
Any smaller time step will result in an increase of CPU time without any corresponding accuracy improvement.
This fine-tuning of the time step should be performed for each orbital regime and for different spacecraft characteristics to ensure that a semi-analytical method gives the best performance.
Similarly, the four additional parameters listed in~\cref{tab:STELA_LEO_params} should also be tuned, which can be time-consuming.
In contrast, the accuracy of a non-averaged method is completely determined by one parameter, that is the integration tolerance.
The optimization of only one parameter in \Thalassa{} (rather than five as in \STELA{}) results in a much simpler propagator configuration.

\subsection{Performance of the non-averaged methods}
The maximum accuracy reached by \Thalassa{} is limited by the accumulation of round-off error (leftmost points in the panels in~\cref{fig:LEO_batch_COE}).
The EDromo and SS element methods are more efficient than Cowell and KS, which are based on coordinates.
For the same computational time, element methods improve the accuracy with respect to \STELA{} by up to three orders of magnitude in each of the orbital elements.
Equivalently, they endow a reduction in CPU time with respect to Cowell by a factor of 3 for the same accuracy.
There is no strong speed advantage in choosing the Kustaanheimo Stiefel formulation over Cowell in the propagation of a quasi-circular orbit, since the rate of change of the physical time is equal to that of the fictitious time, except from a multiplicative constant ($\mathrm{d}t = r \mathrm{d}s$).
Nevertheless, KS does achieve a higher accuracy especially when used in conjunction with a linear time element.
Computing the position along the orbit with an accuracy greater than \SI{1}{\degree} (equivalent to about \SI{120}{\kilo\meter} in the along-track direction) is only possible if regularized formulations are employed.
This is remarkable given the long time span of the integration, which is of about \num{e5} orbital periods.


\subsection{Re-entry date prediction accuracy}
\begin{figure}
\centering
\includegraphics[scale=0.5,clip,trim={0in 0.0in 0in 0.0in}]{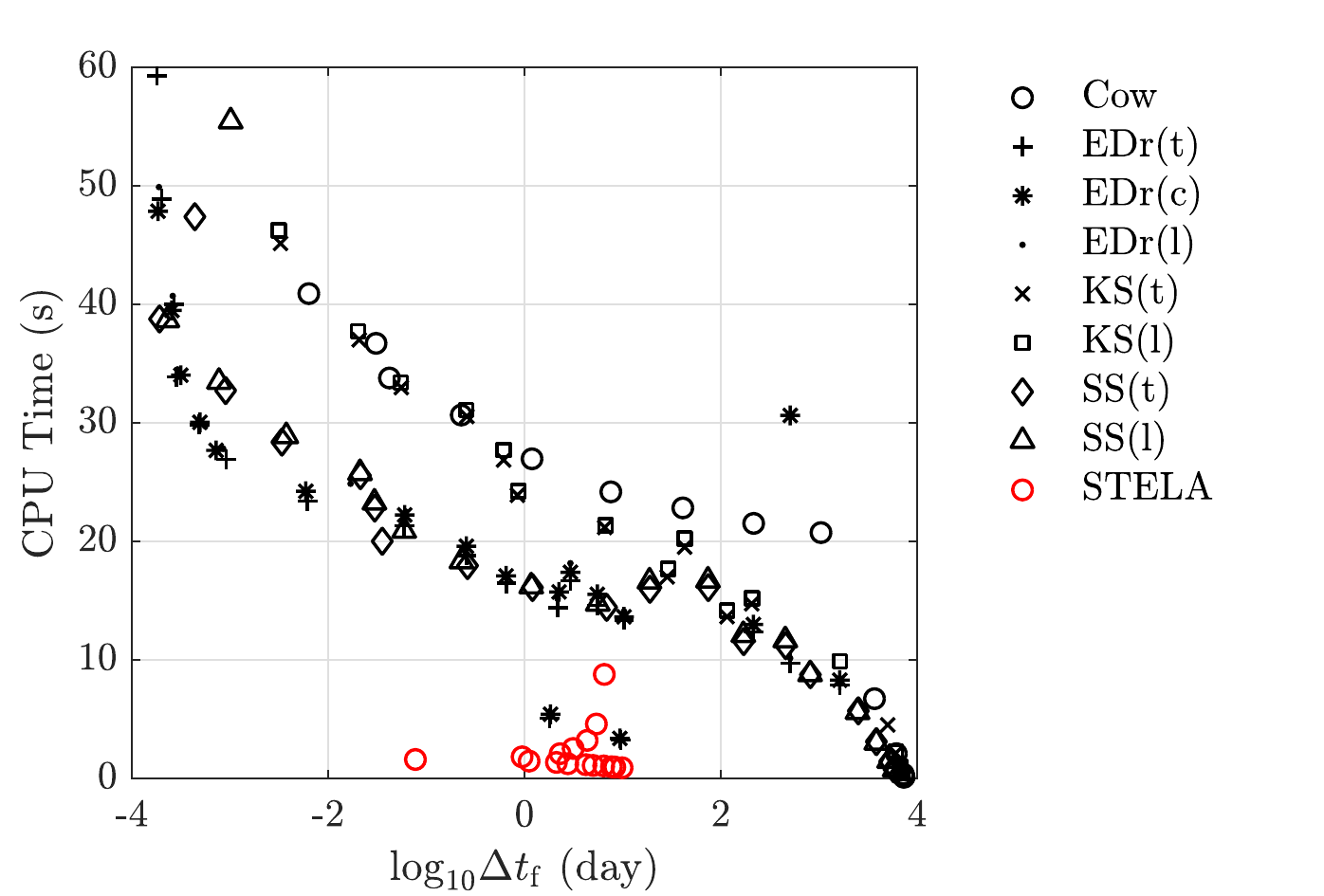}
\caption{Measured CPU time as a function of the errors on the re-entry date for the LEO test case, with respect to a reference solution obtained in quadruple precision. Refer to \cref{fig:LEO_batch_COE} for the description of the numerical tests and of the labels.
\label{fig:LEO_batch_JD}}
\end{figure}
We use a procedure analogous to that of the previous section to compute the error in the re-entry date with respect to the reference solution for \Thalassa{} and \STELA{}. The US76 atmospheric model is used for both of them.
\Cref{fig:LEO_batch_JD} shows that the dynamical and model truncation errors in \STELA{} limit its re-entry date prediction accuracy to \num{1} day after 20 years of propagation.
\STELA{} is three times faster than \Thalassa{} for the same accuracy in re-entry time.
The difference in performance of the methods is similar to the case in which we take the error on the orbital elements as an accuracy metric (compare \cref{fig:LEO_batch_JD} against \cref{fig:LEO_batch_COE}).
Regarding the formulations implemented in \Thalassa{}, EDromo provides the most accurate re-entry predictions.

We also repeated the propagation of the initial conditions in \cref{tab:ICs_LEO}, but changing the initial inclination to the ``critical'' value of \SI{63.4}{\degree}, for which $\dot{\omega} \approx 0$.
All the perturbations considered until now were kept active in the test.
The qualitative behavior of the trajectories and the re-entry dates obtained with both codes were in good agreement.
We highlight that \STELA{} was able to provide reliable estimations of the re-entry epoch even close to the condition $\dot{\omega} \approx 0$, which is an intrinsic singularity of the main problem~\citep{Coffey1986}.
\begin{table}[t]
\centering
\caption{Initial modified Julian date and osculating orbital elements for the GTO test case.\label{tab:ICs_GTO}}
\[
        \begin{array}{cS[tight-spacing=true,table-format=5.7e0]s}
        \toprule
        \text{MJD}            & 57249.958333  &             \\
        \hat{a}               & 24326.18      & \kilo\metre \\
        \hat{e}               & 0.73          &             \\
        \hat{i}               & 10            & \degree     \\
        \hat{\Omega}          & 310           & \degree     \\
        \hat{\omega}          & 0             & \degree     \\
        \hat{M}               & 180           & \degree     \\
        \bottomrule
        \end{array}
        \]
\end{table}
\section{GTO numerical test case}
\label{sec:GTO_test}
The same numerical experiments as in the previous section are repeated for a set of initial conditions representing a GTO orbit, which is displayed in~\cref{tab:ICs_GTO}.
The physical model is unchanged.
The mass, area, and drag coefficient of the spacecraft are set at $M = \SI{1000}{\kilo\gram}$, $A = \SI{10}{\square\metre}$, $C_D = 2.2$, respectively.

\subsection{GTO dynamics and impact on the integration error}
\begin{figure}[t]
        \centering
        \includegraphics[width=0.8\columnwidth]{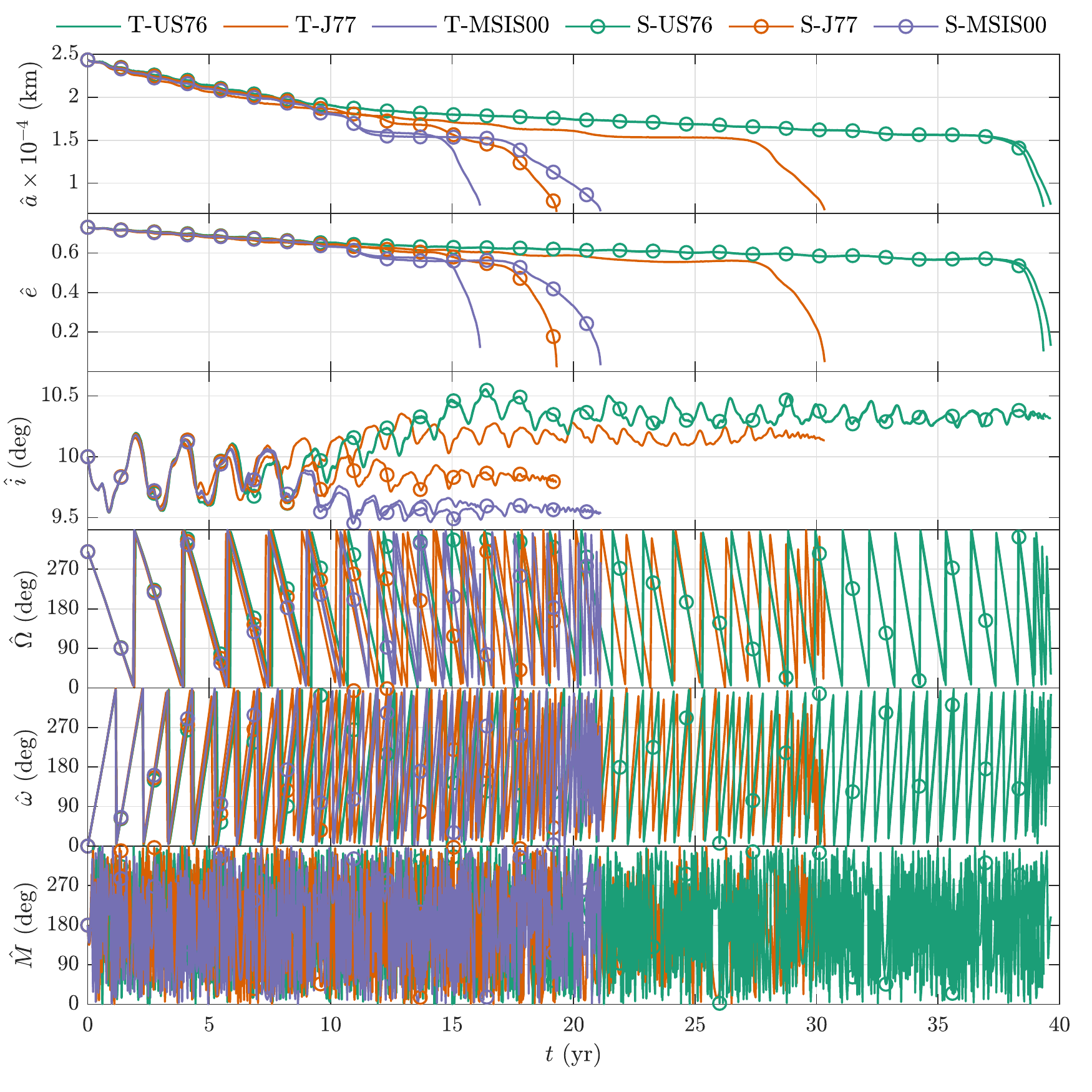}
        \caption{Osculating orbital elements as a function of time for the propagation of the GTO initial conditions in \cref{tab:ICs_GTO} until re-entry.
        Refer to \cref{fig:LEO_COE}\label{fig:GTO_COE} for the interpretation of the curves.} 
\end{figure}
\begin{table}[t]
        \centering
        \caption{Values of the parameters used in the \STELA{} propagations for the GTO test case (curves ``S-US76'', ``S-J77'', and ``S-MSIS00'' in \cref{fig:GTO_COE}). The meaning of the parameters is explained in \cref{sec:STELA_mathmod}. \label{tab:STELA_GTO_params}}
        \[
            \begin{array}{c*5{S[tight-spacing=true]}}
                \toprule
                {\Delta {t}} & {l_\mathrm{M}} & {N_\mathrm{tess}} & N_\mathrm{drag} & {M_\mathrm{quad}} \\
                \midrule
                \SI{6}{\hour} & 8 & 5 & 1 & 67 \\
                \bottomrule
            \end{array}
        \]
\end{table}
The complex dynamics of GTOs is dictated by the interplay between lunisolar perturbations and atmospheric drag, and the evolution of any single orbit can only be predicted in statistical terms.
Nevertheless, it is still possible to predict the evolution of a single orbit over a certain time span if the integration errors are reduced to within a few orders magnitude of the machine zero.
Therefore we build reference solutions for all atmospheric models by propagating the initial conditions in \cref{tab:ICs_GTO} with \Thalassa, with a very strict solver tolerance of \num{e-15}.
For the US76 model, we also propagate in quadruple precision and with a tolerance of \num{e-23}.
The quadruple and double precision solutions are in excellent agreement, indicating that double precision is adequate to integrate this orbit accurately.
Regarding \STELA{}, its parameters were chosen by trial and error as to make the re-entry date converge within an acceptable CPU time, analogously to \cref{sec:LEO_test}.
Note that the integration step (${\Delta {t}}$) of \STELA{} needs to be four times smaller than in the LEO case, notwithstanding the longer GTO orbital period.

\Cref{fig:GTO_COE} shows the time histories of the orbital elements until re-entry for both codes and all atmospheric models.
Differences between the two codes are more pronounced than in the LEO case.
The visible discrepancy in the last 3 years of propagation with the US76 model is due to the accumulation of dynamical error, which may be aggravated in the presence of drag.
The mean rate of change of the elements due to atmospheric drag is obtained in \STELA{} by averaging the related acceleration at $M_\mathrm{quad}$ points on the \emph{osculating} orbit by numerical quadrature.
However, the dynamical and short-periodic errors affecting the osculating elements ultimately lead to an error on the computed altitude and thus on the density and drag acceleration.
This effect can be relevant in the latter part of the propagation, when the dynamical error has accumulated significantly and the orbit crosses the densest layers of the atmosphere.
\begin{figure}[t]
        \centering
        \includegraphics[scale=0.7]{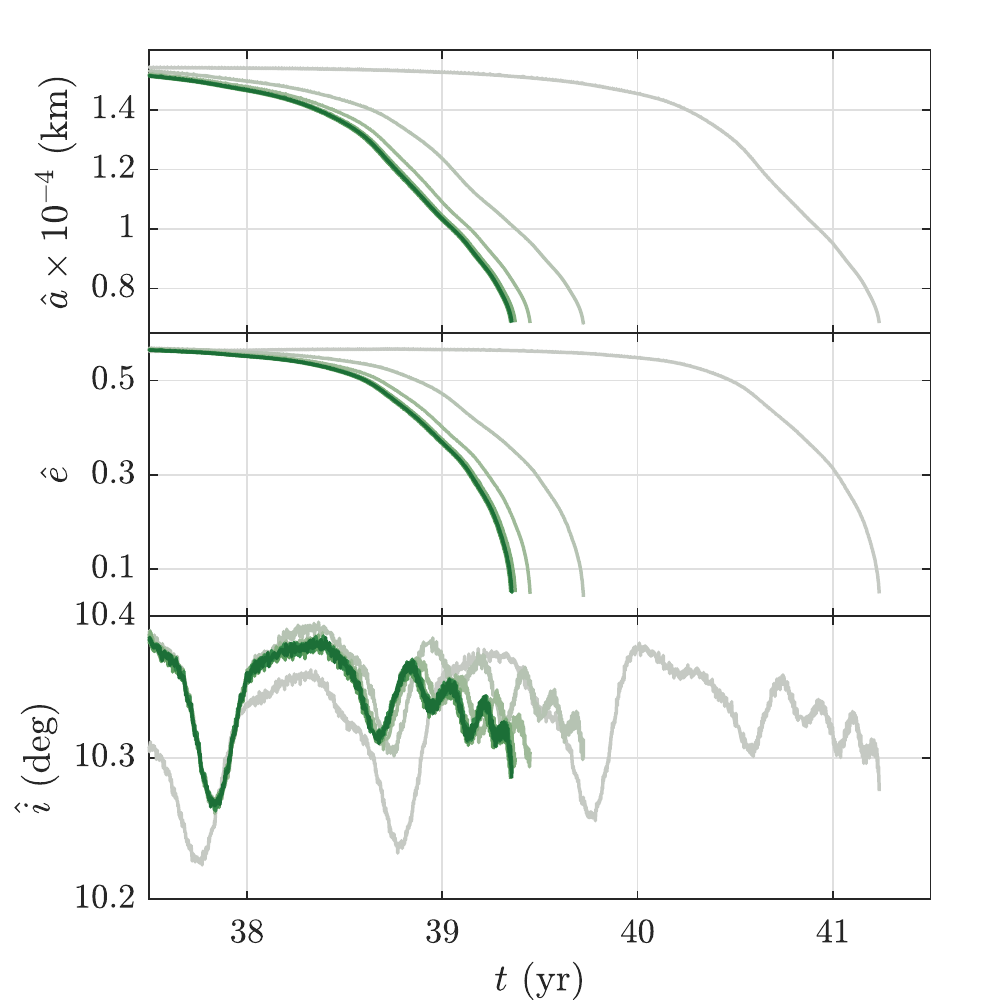}
        \caption{Osculating \STELA{} trajectories for the order of truncation of the lunisolar expansions $l_\mathrm{M}$ varying from \num{2} (gray) to \num{8} (dark green). Only $(\hat{a},\hat{e},\hat{i})$ are shown.\label{fig:GTO_COE_lM}} 
\end{figure}

Due to the high sensitivity of this orbit, a higher truncation order $l_\mathrm{M}$ of the lunisolar perturbing function expansion is needed to achieve a good convergence of the re-entry date.
The impact of $l_\mathrm{M}$ is apparent from \cref{fig:GTO_COE_lM}, which highlights that the evolution of the trajectory in the last years of propagation (the most relevant in understanding the evolution of the re-entry process) changes substantially with $l_\mathrm{M}$.
Choosing an adequate $l_\mathrm{M}$ value increases the effort needed for fine-tuning the semi-analytical propagator, as noted in \cref{sec:LEO_SA_perf}.

\begin{figure}
\centering
\includegraphics[scale=0.5,clip,trim={0in 0.1in 0in 0.6in}]{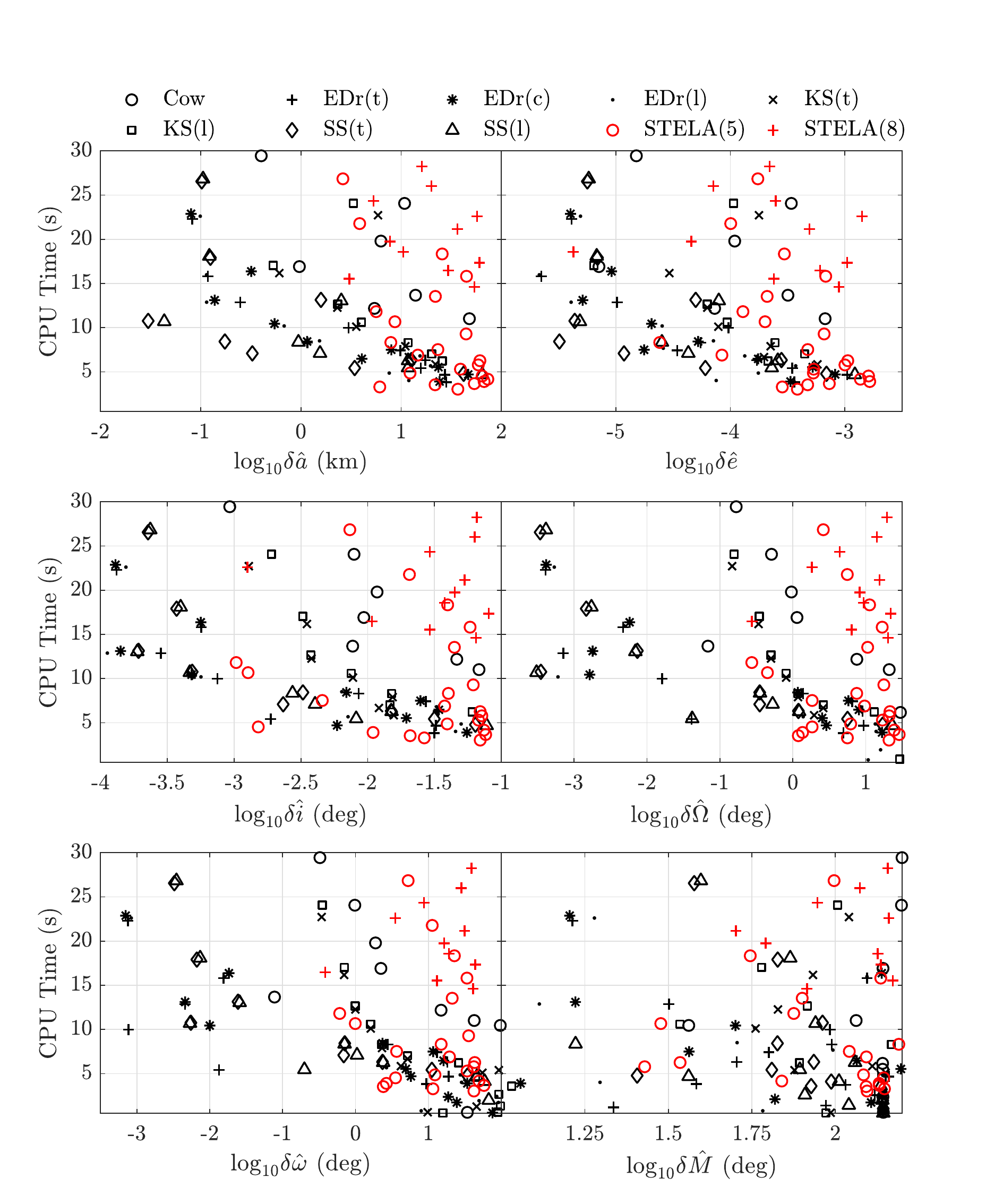}
\caption{Measured CPU time as a function of the errors on the osculating orbital elements for the GTO test case (considering the US76 atmospheric model) with respect to a reference solution computed in quadruple precision. Error values are obtained by changing the solver tolerance (for \Thalassa{}, black markers) or the solver step size (for \STELA{}, red markers), and are all measured after 20 years of propagation of the initial conditions in \cref{tab:ICs_GTO}. Refer to \cref{fig:LEO_batch_COE} for the description of the labels. We report \STELA{} propagations for $l_\mathrm{M} = 5$ and $l_\mathrm{M} = 8$. \label{fig:GTO_batch_COE}}
\end{figure}

\subsection{Performance of the methods}
\Cref{fig:GTO_batch_COE} shows the measured CPU time as a function of the errors on the orbital elements for all methods.
The errors are taken with respect to a reference trajectory computed in quadruple precision in \Thalassa{} (with a solver tolerance of \num{e-23}) for the US76 atmospheric model after 20 years of propagation.
In order to investigate the impact of the model truncation error, we report results obtained with both $l_\mathrm{M} = 5$ and $l_\mathrm{M} = 8$ for \STELA{}.
We vary the \Thalassa{} solver tolerance between \num{e-4} and \num{e-13}, and the \STELA{} time step between \num{0.1} and \num{3} days.

As in the previous case, the regularized formulations implemented in \Thalassa{} reach significantly higher accuracy than the semi-analytical approach in \STELA{} for all orbital elements except $\hat{M}$.
Regularization is highly beneficial for the integration of this test case, as the Cowell formulation has high CPU time at a relatively poor accuracy, and the best performing formulations are the EDromo and SS element methods.
The performance of \Thalassa{} is superior to that of \STELA{} since, by using regularized formulations, it is possible to get a solution that is more accurate than the semi-analytical approach with comparable CPU time.


Increasing $l_\mathrm{M}$ after a certain threshold does not improve the accuracy of the \STELA{} solutions.
Examination of \cref{fig:GTO_batch_COE} suggests that the model truncation error does not have a significant impact for $l_\mathrm{M} > 5$.
Since decreasing the time step does not result in improvements either, the large integration errors produced by \STELA{} are ascribable to the dynamical error $\delta{E}_{i,\mathrm{dyn}}$, as explained in \cref{sec:avg_err_analysis}.
We verified that using \Thalassa{} with the truncated expansion of the lunisolar perturbations (\cref{eq:P1Trunc}) shows negligible qualitative improvements in the trajectory for $l_\mathrm{M} > 6$.

The accuracy of the trajectories computed with \STELA{} is strongly sensitive to the value of the time step, as it can be inferred from the noticeable scattering in the plots.
Regularized formulations show a smoother convergence with decreasing tolerance, and the variations in the integration error are more contained.

Errors in mean anomaly are large for all the methods, preventing the accurate computation of the position vector.
This is not an issue, since the uncertainty embedded in the physical model and in the orbit determination prevents the accurate recovery of the position over such a long time span in practical computations. 

\begin{figure}
\centering
\includegraphics[scale=0.5,clip,trim={0in 0.0in 0in 0.0in}]{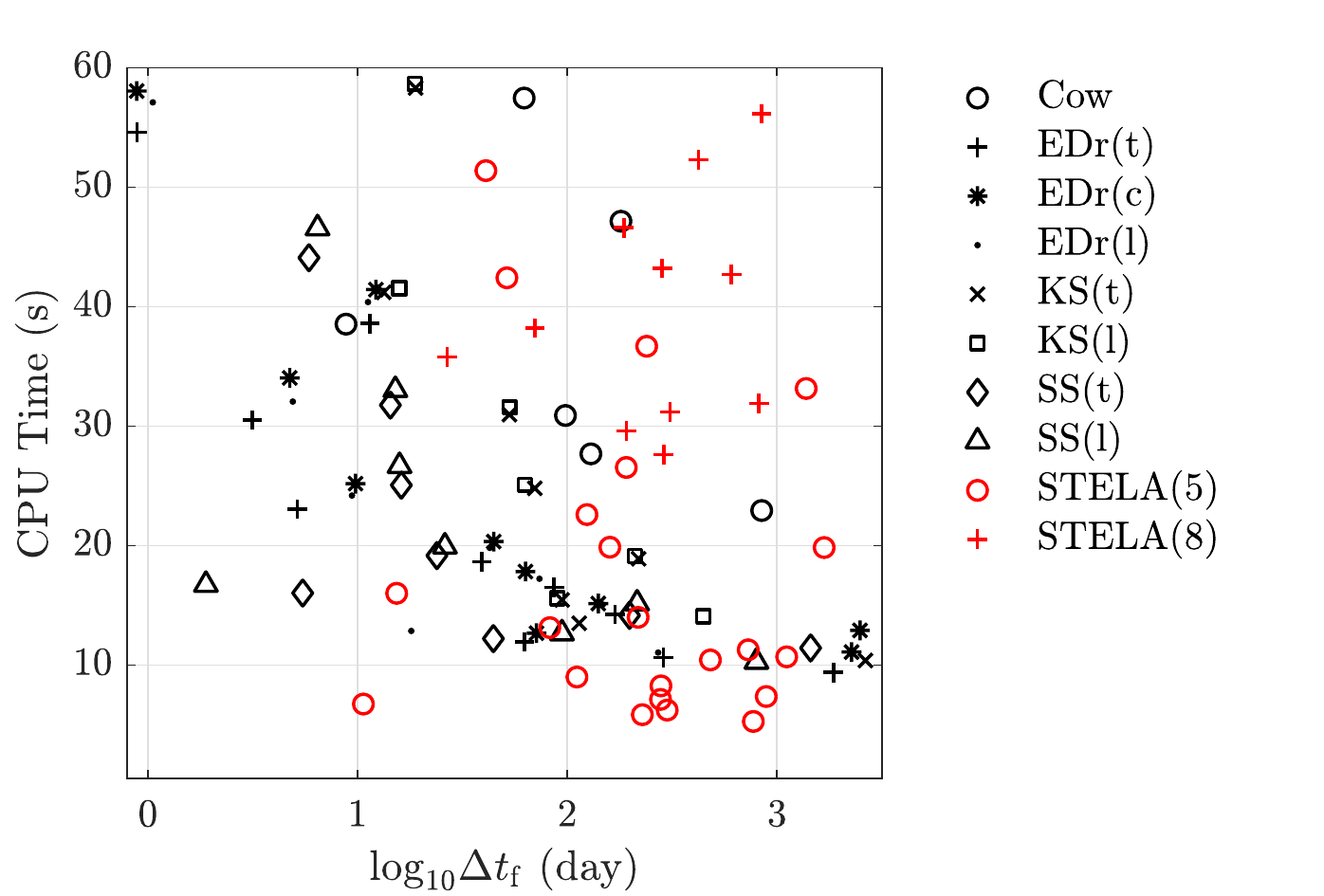}
    \caption{Measured CPU time as a function of the errors on the re-entry date for the GTO test case considering the US76 atmospheric model, with respect to a reference solution. Refer to \cref{fig:LEO_batch_COE} for the description of the numerical tests and of the labels. We report \STELA{} propagations for $l_\mathrm{M} = 5$ and $l_\mathrm{M} = 8$. \label{fig:GTO_batch_JD}}
\end{figure}

All of the above considerations also affect the errors on the re-entry date, which are displayed in~\cref{fig:GTO_batch_JD} for all the methods.
In particular, it is possible to constrain the error on the re-entry date to under 10 days (with CPU times between \num{15} and \num{60} seconds) only by using regularized element methods.
Both \STELA{} and the Cowell formulation achieve errors on the re-entry date in the order of 100 days with CPU times of about 20 to 30 seconds.

\section{HEO numerical test case}
\label{sec:HEO_test}
\begin{table}[t]
    \centering
    \caption{Initial modified Julian date and osculating orbital elements for the HEO test case.\label{tab:ICs_HEO}}
    \[
        \begin{array}{cS[tight-spacing=true,table-format=6.8e0]s}
            \toprule
            \text{MJD}            & 56664.86336805  &             \\
            \hat{a}               & 106247.136454   & \kilo\metre \\
            \hat{e}               & 0.75173         &             \\
            \hat{i}               & 5.2789          & \degree     \\
            \hat{\Omega}          & 49.351          & \degree     \\
            \hat{\omega}          & 180             & \degree     \\
            \hat{M}               & 0               & \degree     \\
            \bottomrule
        \end{array}
        \]
\end{table}
We consider the orbit of the proposed Simbol-X mission as a test case representative of a high-altitude HEO.
The initial conditions in \cref{tab:ICs_HEO} were used to benchmark the performance of the semi-analytical propagator by~\citet{Lara2017}, and for the study performed on \Thalassa{} in~\citet{Amato2018}.
The large values of eccentricity and semi-major axis make this orbit a challenging test case for both semi-analytical and numerical methods.
With a period of approximately $\num{4}$ days, the orbit is also close to a $\num{7}:\num{1}$ mean motion resonance with the Moon.
We consider the same physical model as in the previous sections, and a spacecraft mass $M = \SI{1470}{\kilo\gram}$, cross-sectional area $A = \SI{15}{\square\meter}$, and drag coefficient $C_D = \num{2.2}$.
Even if the semi-major axis is very large, the high eccentricity of this orbit causes the spacecraft to cross the atmosphere in some parts of the trajectory.
Thus we leave the atmospheric drag perturbation active, with the air density computed through the US76 model only.

\subsection{Impact of dynamical and model truncation errors}
\begin{table}[t]
    \centering
    \caption{Values of the parameters used in the \STELA{} propagations for the HEO test case (curves ``STELA, $l_\mathrm{M}=5$'' and ``STELA, $l_\mathrm{M}=8$'' in \cref{fig:HEO_COE}). The meaning of the parameters is explained in \cref{sec:STELA_mathmod}. \label{tab:STELA_HEO_params}}
    \[
        \begin{array}{c*5{S[tight-spacing=true]}}
            \toprule
            {\Delta {t}} & {l_\mathrm{M}} & {N_\mathrm{tess}} & N_\mathrm{drag} & {M_\mathrm{quad}} \\
            \midrule
            \SI{48}{\hour} & \text{5 and 8} & 2.5 & 1 & 33 \\
            \bottomrule
        \end{array}
    \]
\end{table}
\begin{figure}[t]
    \centering
    \includegraphics[width=0.8\columnwidth,clip,trim={0.15in 0.1in 0.15in 0.05in}]{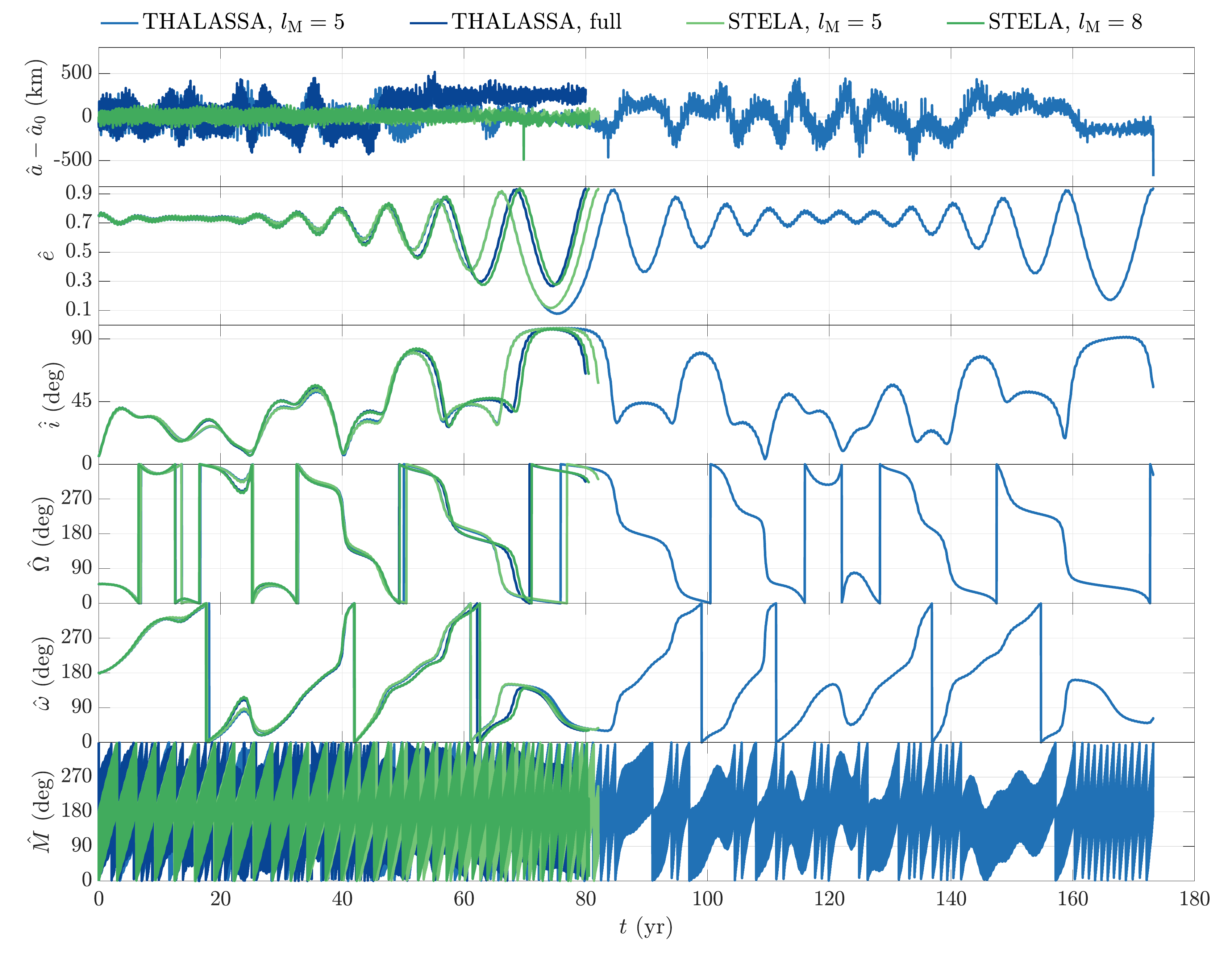}
    \caption{Osculating orbital elements as a function of time for the propagation of the HEO initial conditions in \cref{tab:ICs_HEO} until re-entry. Blue curves refer to propagations with \Thalassa{} considering either the full expression for lunisolar perturbations (\cref{eq:P1Cart}) or the one truncated at $l_\mathrm{M} = 5$ (\cref{eq:P1Trunc}). The re-entry epoch for the latter solution is delayed by about 90 years with respect to the others. Green curves refer to propagations with \STELA{} for $l_\mathrm{M} = 5$ and $l_\mathrm{M} = 8$.\label{fig:HEO_COE}} 
\end{figure}

Trajectories resulting from the propagation of the initial conditions in~\cref{tab:ICs_HEO} are presented in~\cref{fig:HEO_COE}.
To investigate the impact of the model truncation error, we show trajectories obtained with \Thalassa{} expressing third-body perturbations through either \cref{eq:P1Cart} (i.e. the ``full'' expression), or \cref{eq:P1Trunc} with $l_\mathrm{M} = 5$ for both the Sun and the Moon.
Both trajectories are computed with strict solver tolerances in order for the numerical error to have a negligible impact, and we take the one with the ``full'' expression as the reference trajectory.
We use the \STELA{} parameters in~\cref{tab:STELA_HEO_params}, which have been found through the same trial-and-error procedure described in the previous sections.
All the propagations are stopped at re-entry, which takes place due to the increase in eccentricity caused by lunisolar perturbations.
The re-entry epoch changes considerably according to the different solutions.

The impact of $l_\mathrm{M}$ on the evolution of the orbital elements is significant, implying that the model truncation error is important for this case.
The \STELA{} and \Thalassa{} solutions for $l_\mathrm{M} = 5$ are qualitatively different from the \Thalassa{} reference, as is particularly evident from the plots of eccentricity and inclination.
The frequency of the oscillations of these elements is affected by substantial errors for $l_\mathrm{M} = 5$.
Since re-entry can only take place when the eccentricity reaches a peak in the long-periodic oscillations, small errors in their amplitude caused by the model truncation lead to an overestimation of the re-entry epoch.
Also for $l_\mathrm{M} = 5$, \STELA{} reports a re-entry at about \num{81} years from the initial epoch, while in the \Thalassa{} solution with $l_\mathrm{M} = 5$ re-entry takes place \num{93} years later.
\STELA{} considers re-entry to take place when the osculating radius of perigee is less than \SI{120}{\kilo\meter}, while \Thalassa{} checks this condition on the osculating radius $\hat{r}$.
This leads to an error in the lifetime estimation of \num{92} years, since the condition on the instantaneous radius of perigee does not correspond to a physical re-entry.
For $l_\mathrm{M} < 5$, discrepancies with respect to the reference solutions are even more relevant; however, we omit the corresponding trajectories here as the interested reader can find these results in~\citet{Amato2018}.

The dynamical error is of minor importance, but it can explain the remaining discrepancies between \STELA{} and \Thalassa{} solutions.
Considering the orbital elements of the Moon as constant during the averaging operation generates a non-negligible dynamical error, since the Moon moves on an arc of approximately \SI{53}{\degree} along its orbit during a single orbital period of the spacecraft.
Moreover, the proximity to the $7:1$ mean motion resonance may cause long-periodic effects to be neglected within the single averaging approach.
In fact, this is a possible explanation to the sudden divergence of the \STELA{} and \Thalassa{} solutions for $l_\mathrm{M} = 5$ at 75 years from the starting epoch.

Ultimately, the accumulation of model truncation and dynamical errors are ascribable to the same cause, which is the large value of the ratio between the mean semi-major axes of the orbiter and of the Moon, $(a/a_\leftmoon) \approx 0.3$.
This implies that the frequencies pertaining to the orbital motion of the spacecraft and to the perturbing body are not well separated, which reduces the efficiency of the semi-analytical approach.

\subsection{Performance of the methods}
\begin{figure}
    \centering
    \includegraphics[scale=0.5,clip,trim={0in 0.15in 0in 0.6in}]{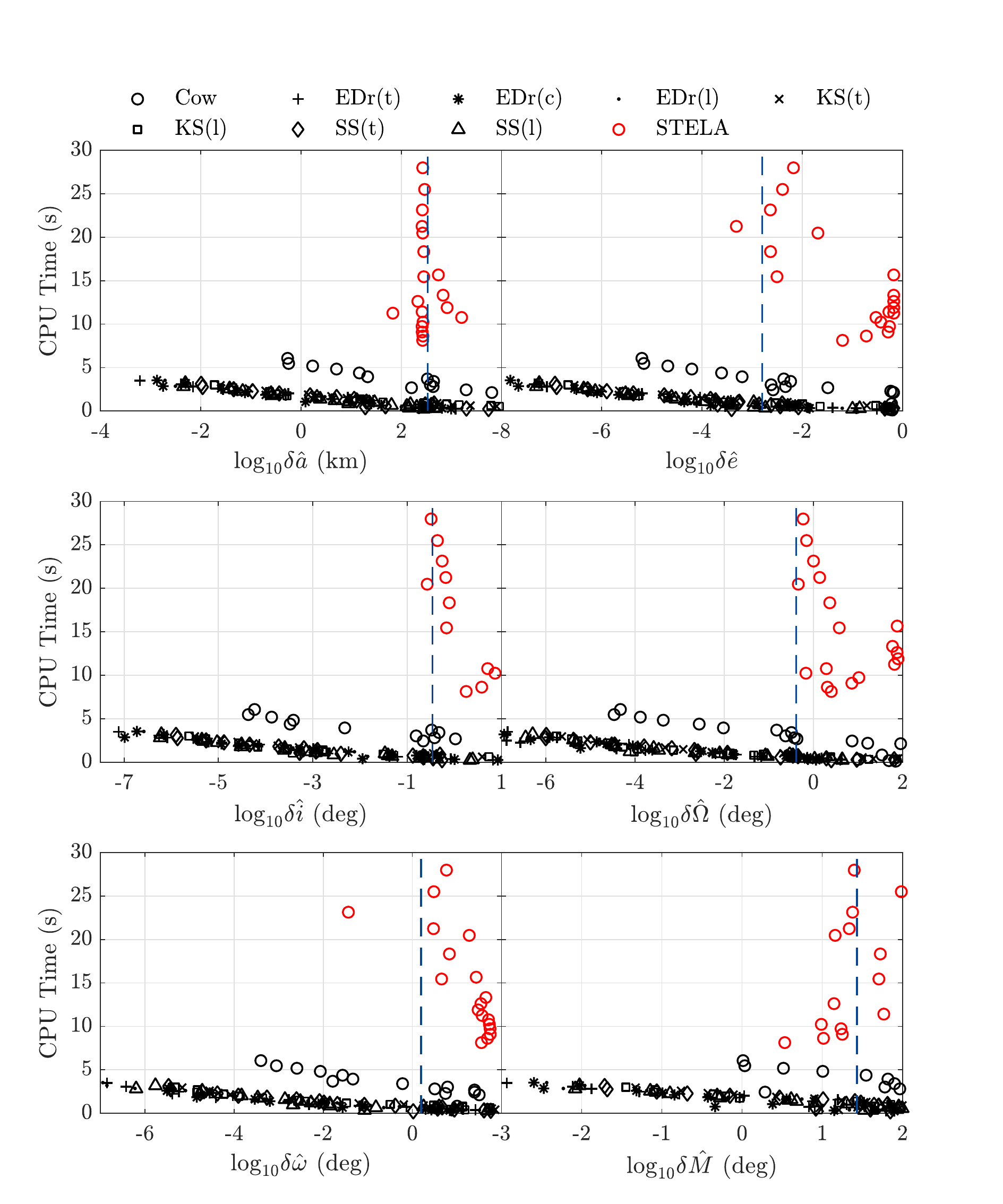}
    \caption{Measured CPU time as a function of the errors on the osculating orbital elements for the HEO test case with respect to a reference solution computed in quadruple precision. Error values are obtained by changing the solver tolerance (for \Thalassa{}, black markers) or the solver step size (for \STELA{}, red markers), and are all measured after 75 years of propagation of the initial conditions in \cref{tab:ICs_HEO}. The dashed blue line denotes the value of the model truncation error at the end of the propagation. Refer to \cref{fig:LEO_batch_COE} for the description of the labels. We report \STELA{} propagations for $l_\mathrm{M} = 8$. \label{fig:HEO_batch_COE}}
\end{figure}

\begin{figure}
    \centering
    \includegraphics[scale=0.5,clip,trim={0in 0.1in 0in 0.6in}]{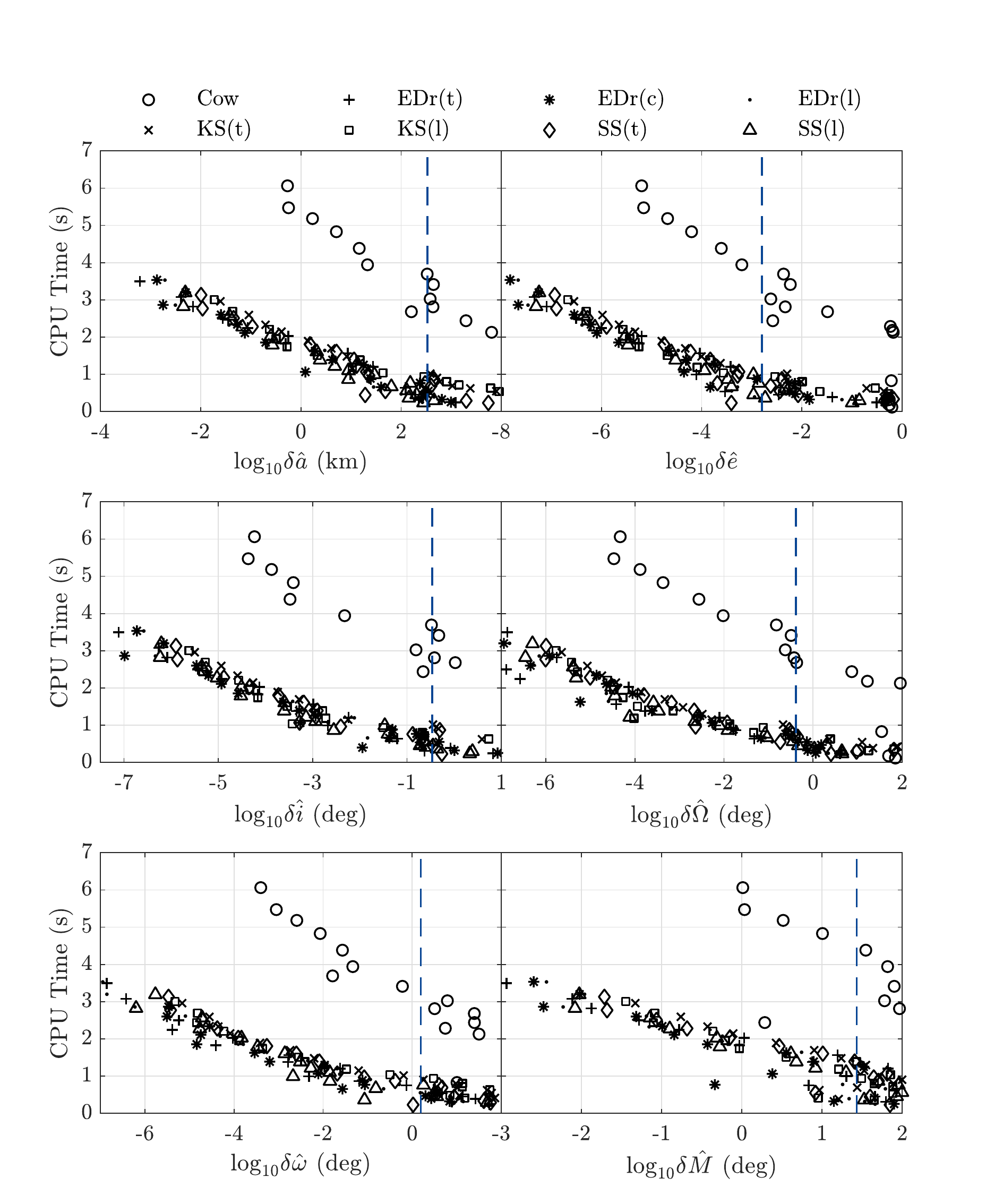}
    \caption{Zoom-in of~\cref{fig:HEO_batch_COE} to highlight the performance of the formulations implemented in \Thalassa{}. \label{fig:HEO_batch_COE_Zoom}}
\end{figure}

The CPU time as a function of the errors on the osculating orbital elements is shown in \cref{fig:HEO_batch_COE,fig:HEO_batch_COE_Zoom}, where all the propagations are stopped at 75 years after the initial epoch.
For \Thalassa{}, we consider the same range of solver tolerance as in the previous sections, while the \STELA{} time step is between \num{5} and \num{20} days.
We also generate a \Thalassa{} solution in quadruple precision with a solver tolerance of \num{e-18}, and using the expansion ${\bm{P}'}^\mathrm{T}$ truncated at $l_\mathrm{M} = 8$, which is the same order that we take into account for the \STELA{} propagations.
Since \Thalassa{} uses non-averaged equations the solution is free from both dynamical error and error on short-periodic terms, and due to the employment of quadruple precision and a very strict tolerance it can be considered free from both round-off and numerical truncation error. 
Thus, the total integration error for this solution, which is represented in the panels in \cref{fig:HEO_batch_COE} by dashed blue lines, will be only due to the model truncation error contribution,
\[
{\delta}\hat{E}_i = {\delta}E_{i,\mathrm{mod}}.
\]

\STELA{} exhibits very coarse accuracy due to the aforementioned accumulation of mean integration error, with \Thalassa{} being six or more orders of magnitude more accurate on all the orbital elements.
The total integration error is dominated by the contribution of model truncation for all the elements except the semi-major axis: its mean value is invariant in the single-averaged restricted three-body problem, which is a good approximation of the physical model for this test if the Moon is considered as the secondary mass.   Decreasing the value of the time step makes the \STELA{} error values converge to the total integration error of the truncated \Thalassa{} solution, confirming the importance of the model truncation error.
Note that carrying the third-body expansion to such a high order substantially increases the computational time. As a consequence, \STELA{} is ten times slower than \Thalassa{} to achieve the same accuracy.

\Cref{fig:HEO_batch_COE_Zoom} is a zoom of the lower region of the panels in~\cref{fig:HEO_batch_COE}, which allows to examine more in detail the performance of the \Thalassa{} formulations.
Since the trajectory is highly perturbed by the Sun and the Moon, there is no particular advantage in using element-based methods.
However, regularization is highly beneficial to the integration, as all the regularized formulations display CPU times of less than half {that of the unregularized} Cowell for the same accuracy.

\section{Conclusions}
\label{sec:conclu}
This paper presents a study of semi-analytical and non-averaged orbit propagation methods for Earth satellites.
By {analyzing} the approximations involved in the method of averaging, we break down the total integration error of a semi-analytical method with respect to an osculating reference (or ``true'') solution into components with distinct mathematical causes.
These are the \emph{dynamical error}, originating in the approximations involved in performing the averaging integrals, the \emph{model truncation error}, due to the truncation of the expansions of the disturbing functions, and the error involved in the calculation of \emph{short-periodic terms}.
All of the above contributions to the total integration error do not depend on the numerical integration method, and can only be abated by resorting to more refined analytical developments.
In contrast, the \emph{numerical error}, which is the last component of the total integration error, can be significantly mitigated with appropriately configured numerical methods.
In this sense, the accuracy of semi-analytical methods is intrinsically limited with respect to the non-averaged.
Moreover, semi-analytical methods involve several parameters that have to be tuned (often by trial and error) to increase performance, while non-averaged methods only require choosing the solver step size or local truncation error tolerance.

We implemented a collection of non-averaged methods in the \Thalassa{} orbit propagation code, consisting of regularized formulations of the equations of motion and of the Cowell formulation (i.e., the unregularized equations in rectangular coordinates).
Their performance is compared to that of the semi-analytical method implemented in the \STELA{} software.
The physical model implemented in \Thalassa{} was aligned to machine precision with that implemented in \STELA{}.
In this way, any discrepancy between \Thalassa{} and \STELA{} solutions can be ascribed to the error components mentioned above.
\Thalassa{} also makes use of a highly efficient and non-singular algorithm for the calculation of the perturbing part of the geopotential.

We presented results from numerical test cases involving the propagation over several decades of initial conditions corresponding to a LEO, a GTO, and a high-altitude HEO.
In the LEO case, the formulations implemented in \Thalassa{} require twice the computational time as \STELA{} to attain comparable accuracy.
The orbit is quasi-circular and has a small semi-major axis, hence the expansions of the perturbing functions  and of the averaging integrals converge rapidly and the semi-analytical method is very efficient.
Regularized formulations achieve errors smaller by three orders of magnitude at the expense of the computational cost; among these, element-based methods are particularly efficient.
On the other hand, regularization is highly beneficial for the GTO orbit, as regularized formulations endow an increase in accuracy of up to two orders of magnitude with respect to \STELA{}.
Dynamical and short-periodic error contributions are particularly important for the GTO, making semi-analytical methods less advantageous than non-averaged methods based on regularized formulations.
For the HEO orbit, the large values of semi-major axis and eccentricity imply that the model truncation error is very significant, and the lunar perturbing function expansion in \STELA{} must be carried out to the highest order (the eighth) to retain qualitative similarity between its solution and the reference.
We offer a quantitative proof by building an additional solution with \Thalassa{} in which the third-body perturbing acceleration is explicitly written as the gradient of the \emph{truncated} perturbing function through a novel expression.
For the HEO, \Thalassa{} is considerably more {efficient: by} using regularized element methods the computational time is abated tenfold with respect to \STELA{} for the same accuracy.
When choosing small values of the solver tolerance, the accuracy reached by \Thalassa{} is significantly higher than that of \STELA{}.

We propose to expand the present work by investigating the performance of the codes in terms of function evaluations, rather than CPU time, and by comparing results of Monte Carlo runs for the estimation of GTO lifetimes.
Additionally, the osculating trajectories produced by \Thalassa{} could be numerically averaged after the propagation in order to compare them to the mean trajectories produced by \STELA{}.

\begin{acknowledgements}
D. A. gratefully acknowledges the advice by Juan-Félix San Juan, Martin Lara, Denis Hautesserress, and Florent Deleflie during numerous occasions, and in particular during the KePASSA conferences and CNES conference on HEO orbits.
D.A. recognizes the extremely helpful assistance by Hodei Urrutxua in the implementation of the non-singular geopotential code.
\end{acknowledgements}

\section*{Conflict of interest}
The authors declare that they have no conflict of interest.

\bibliographystyle{spbasic}      
\bibliography{references}   

\begin{thebibliography}{63}
\providecommand{\natexlab}[1]{#1}
\providecommand{\url}[1]{{#1}}
\providecommand{\urlprefix}{URL }
\expandafter\ifx\csname urlstyle\endcsname\relax
  \providecommand{\doi}[1]{DOI~\discretionary{}{}{}#1}\else
  \providecommand{\doi}{DOI~\discretionary{}{}{}\begingroup
  \urlstyle{rm}\Url}\fi
\providecommand{\eprint}[2][]{\url{#2}}

\bibitem[{Amato et~al(2017)Amato, Baù, and Bombardelli}]{Amato2017}
Amato D, Baù G, Bombardelli C (2017) Accurate orbit propagation in the
  presence of planetary close encounters. Mon Not R Astron Soc
  470(2):2079--2099, \doi{10.1093/mnras/stx1254}

\bibitem[{Amato et~al(2018)Amato, Rosengren, and Bombardelli}]{Amato2018}
Amato D, Rosengren AJ, Bombardelli C (2018) {THALASSA}: a fast orbit propagator
  for near-earth and cislunar space. In: 2018 Space Flight Mechanics Meeting,
  American Institute of Aeronautics and Astronautics, \doi{10.2514/6.2018-1970}

\bibitem[{Anselmo et~al(1997)Anselmo, Cordelli, Farinella, Pardini, and
  Rossi}]{Anselmo1997}
Anselmo L, Cordelli A, Farinella P, Pardini C, Rossi A (1997) Modelling the
  evolution of the space debris population: Recent research work in {Pisa}. In:
  Second European Conference on Space Debris, Darmstadt, Germany, ESA Special
  Publication, vol 393, pp 339--344,
  \urlprefix\url{http://adsabs.harvard.edu/abs/1997ESASP.393..339A}

\bibitem[{{Battin}(1999)}]{Battin1999}
{Battin} RH (1999) {An Introduction to the Mathematics and Methods of
  Astrodynamics}, revised edn. AIAA Education Series, American Institute of
  Aeronautics and Astronautics, Reston, VA, USA, \doi{10.2514/4.861543}

\bibitem[{{Ba\`{u}} et~al(2014){Ba\`{u}}, {Urrutxua}, and
  {Pel{\'a}ez}}]{Bau2014a}
{Ba\`{u}} G, {Urrutxua} H, {Pel{\'a}ez} J (2014) {EDromo}: an accurate
  propagator for elliptical orbits in the perturbed two-body problem. Adv
  Astronaut Sci 152:379--399, proceedings of the 24th AAS/AIAA Space Flight
  Mechanics Meeting, January 26-30, 2014, Santa Fe, New Mexico

\bibitem[{Baù and Bombardelli(2014)}]{Bau2014}
Baù G, Bombardelli C (2014) Time elements for enhanced performance of the
  {Dromo} orbit propagator. Astronom J 148(3):43,
  \doi{10.1088/0004-6256/148/3/43}

\bibitem[{Baù et~al(2015)Baù, Bombardelli, Peláez, and Lorenzini}]{Bau2015}
Baù G, Bombardelli C, Peláez J, Lorenzini E (2015) Non-singular orbital
  elements for special perturbations in the two-body problem. Mon Not R Astron
  Soc 454:2890--2908, \doi{10.1093/mnras/stv2106},
  \urlprefix\url{http://adsabs.harvard.edu/abs/2015MNRAS.454.2890B}

\bibitem[{Biancale et~al(2000)Biancale, Balmino, Lemoine, Marty, Moynot,
  Barlier, Exertier, Laurain, Gegout, Schwintzer, Reigber, Bode, König,
  Massmann, Raimondo, Schmidt, and Yuan~Zhu}]{Biancale2000}
Biancale R, Balmino G, Lemoine JM, Marty JC, Moynot B, Barlier F, Exertier P,
  Laurain O, Gegout P, Schwintzer P, Reigber C, Bode A, König R, Massmann FH,
  Raimondo JC, Schmidt R, Yuan~Zhu S (2000) A new global earth's gravity field
  model from satellite orbit perturbations: {GRIM}5-s1. Geophys Res Lett
  27(22):3611--3614, \doi{10.1029/2000gl011721}

\bibitem[{Bond and Allman(1996)}]{Bond1996}
Bond VR, Allman MC (1996) Modern Astrodynamics, Princeton University Press,
  Princeton, NJ, United States, pp 117--146

\bibitem[{{Broucke}(1966)}]{Broucke1966}
{Broucke} R (1966) {Regularized special perturbation techniques using
  Levi-Civita variables}. In: 3rd and 4th Aerospace Sciences Meeting, American
  Institute of Aeronautics and Astronautics, paper 66-8

\bibitem[{{Brouwer}(1937)}]{Brouwer1937}
{Brouwer} D (1937) {On the accumulation of errors in numerical integration}.
  Astronom J 46:149--153, \doi{10.1086/105423}

\bibitem[{{Bull} et~al(2001){Bull}, {Smith}, {Pottage}, and
  {Freeman}}]{Bull2001}
{Bull} JM, {Smith} LA, {Pottage} L, {Freeman} R (2001) {Benchmarking Java
  against C and Fortran for scientific applications}. In: {Proceedings of the
  2001 joint ACM-ISCOPE conference on Java Grande}, Association for Computing
  Machinery, Palo Alto, CA, USA, pp 97--105

\bibitem[{{Burdet}(1969)}]{Burdet1969}
{Burdet} CA (1969) {Le mouvement {K}eplerien et les oscillateurs harmoniques}.
  J Reine Angew Math 238:71--84

\bibitem[{{CNES}(2016)}]{STELA_Man2016}
{CNES} (2016) {STELA 3.1.1 User Manual}. Tech. rep.,
  \urlprefix\url{https://logiciels.cnes.fr/sites/default/files/Stela-User-Manual_4.pdf}

\bibitem[{Coffey et~al(1986)Coffey, Deprit, and Miller}]{Coffey1986}
Coffey SL, Deprit A, Miller BR (1986) The critical inclination in artificial
  satellite theory. Cel Mech 39(4):365--406, \doi{10.1007/bf01230483}

\bibitem[{Coffey et~al(1998)Coffey, Neal, Visel, and Conolly}]{Coffey1998}
Coffey SL, Neal HL, Visel CL, Conolly P (1998) Demonstration of a
  special-perturbations-based catalog in the naval space command system. Adv
  Astronaut Sci 99(1):227--247

\bibitem[{Dahlquist and Björck(1974)}]{Dahlquist1974}
Dahlquist G, Björck A (1974) Numerical Methods, Dover Publications, Mineola,
  NY, United States, pp 1--20

\bibitem[{{Danielson} et~al(1995){Danielson}, {Sagovac}, {Neta}, and
  {Early}}]{Danielson1995}
{Danielson} DA, {Sagovac} CP, {Neta} B, {Early} LW (1995) {Semianalytic
  Satellite Theory}. Tech. Rep. ADA276836, Department of Mathematics, Naval
  Postgraduate School, Monterey, CA, USA

\bibitem[{Deprit(1975)}]{Deprit1975}
Deprit A (1975) Ideal elements for perturbed keplerian motions. J Res Nat Bur
  Stand 79:1--15,
  \urlprefix\url{http://adsabs.harvard.edu/abs/1975JRNBS..79....1D}

\bibitem[{Dichmann et~al(2013)Dichmann, Lebois, and Carrico}]{Dichmann2013}
Dichmann DJ, Lebois R, Carrico JP (2013) Dynamics of orbits near 3:1 resonance
  in the {Earth-Moon} system. J Astronaut Sci 60(1):51--86,
  \doi{10.1007/s40295-014-0009-x}

\bibitem[{Eichhorn et~al(2017)Eichhorn, Cano, McLean, and
  Anderl}]{Eichhorn2017}
Eichhorn H, Cano JL, McLean F, Anderl R (2017) A comparative study of
  programming languages for next-generation astrodynamics systems. {CEAS} Space
  J 10(1):115--123, \doi{10.1007/s12567-017-0170-8}

\bibitem[{Ely(2014)}]{Ely2014}
Ely TA (2014) Mean element propagations using numerical averaging. J Astronaut
  Sci 61(3):275--304, \doi{10.1007/s40295-014-0020-2}

\bibitem[{{Ferr{\'a}ndiz}(1988)}]{Ferrandiz1988}
{Ferr{\'a}ndiz} JM (1988) {A general canonical transformation increasing the
  number of variables with application in the two-body problem}. Cel Mech
  41:343--357

\bibitem[{{Giacaglia}(1977)}]{Giacaglia1977}
{Giacaglia} GEO (1977) {The equations of motion of an artificial satellite in
  nonsingular variables}. Cel Mech 15:191--215, \doi{10.1007/BF01228462}

\bibitem[{Gkolias et~al(2016)Gkolias, Daquin, Gachet, and
  Rosengren}]{Gkolias2016}
Gkolias I, Daquin J, Gachet F, Rosengren AJ (2016) From order to chaos in
  {Earth} satellite orbits. Astron J 152(5):119,
  \doi{10.3847/0004-6256/152/5/119}

\bibitem[{Golikov(2012)}]{Golikov2012}
Golikov AR (2012) {THEONA}{\textemdash}a numerical-analytical theory of motion
  of artificial satellites of celestial bodies. Cosm Res 50(6):449--458,
  \doi{10.1134/s0010952512060020}

\bibitem[{van~der Ha(1986)}]{vanderHa1986}
van~der Ha JC (1986) Long-term evolution of near-geostationary orbits. J Guid
  Control Dyn 9(3):363--370, \doi{10.2514/3.20115}

\bibitem[{Hoots et~al(2004)Hoots, Schumacher~Jr., and Glover}]{Hoots2004}
Hoots FR, Schumacher~Jr PW, Glover RA (2004) History of analytical orbit
  modeling in the us space surveillance system. J Guid Control Dyn
  27(2):174--185, \doi{10.2514/1.9161}

\bibitem[{{Jacchia}(1977)}]{Jacchia1977}
{Jacchia} LG (1977) {Thermospheric Temperature, Density, and Composition: New
  Models}. SAO Special Report 375

\bibitem[{{Kaula}(1966)}]{Kaula1966}
{Kaula} WM (1966) {Theory of satellite geodesy. Applications of satellites to
  geodesy}. Blaisdell Publishing, Waltham, MA, USA

\bibitem[{Klinkrad et~al(2006)Klinkrad, Martin, and Walker}]{Klinkrad2006}
Klinkrad H, Martin C, Walker R (2006) Space Debris Models and Risk Analysis,
  Springer-Verlag and Praxis Publishing. \doi{10.1007/3-540-37674-7\_5}

\bibitem[{{Kustaanheimo} and {Stiefel}(1965)}]{KS1965}
{Kustaanheimo} P, {Stiefel} EL (1965) {Perturbation theory of {K}epler motion
  based on spinor regularization}. J Reine Angew Math 218:204--219

\bibitem[{Kwok(1986)}]{Kwok1986}
Kwok JH (1986) {The Long-Term Orbit Predictor (LOP)}. Report EM 312/86-151, Jet
  Propulsion Laboratory, Pasadena, CA, United States

\bibitem[{Lamy et~al(2011)Lamy, Le~Fevre, and Sarli}]{Lamy2011}
Lamy A, Le~Fevre C, Sarli B (2011) Analysis of geostationary transfer orbit
  long term evolution and lifetime. In: Proceedings of the
  22\textsuperscript{nd} International Symposium on Space Flight Dynamics, San
  José dos Campos, Brazil

\bibitem[{Lara(2017)}]{Lara2017a}
Lara M (2017) Note on the ideal frame formulation. Cel Mech Dyn Astr
  129:137--151, \doi{10.1007/s10569-017-9770-z},
  \urlprefix\url{http://adsabs.harvard.edu/abs/2017CeMDA.129..137L},
  \eprint{1612.08367}

\bibitem[{Lara et~al(2012)Lara, San-Juan, L{\'{o}}pez, and Cefola}]{Lara2012}
Lara M, San-Juan JF, L{\'{o}}pez LM, Cefola PJ (2012) On the third-body
  perturbations of high-altitude orbits. Cel Mech Dyn Astr 113(4):435--452,
  \doi{10.1007/s10569-012-9433-z}

\bibitem[{{Lara} et~al(2013){Lara}, {San-Juan}, and
  {L{\'o}pez-Ochoa}}]{Lara2013}
{Lara} M, {San-Juan} JF, {L{\'o}pez-Ochoa} LM (2013) {Proper averaging via
  parallax elimination}. Adv Astronaut Sci 150:315--331, proceedings of the
  AAS/AIAA Astrodynamics Specialist Conference, August 2013, Hilton Head, SC,
  USA

\bibitem[{{Lara} et~al(2017){Lara}, {San-Juan}, and {Hautesserres}}]{Lara2017}
{Lara} M, {San-Juan} JF, {Hautesserres} D (2017) {HEOSAT}: a mean elements
  orbit propagator program for highly elliptical orbits. CEAS Space J
  \doi{10.1007/s12567-017-0152-x}

\bibitem[{Le~F{\`{e}}vre et~al(2014)Le~F{\`{e}}vre, Fraysse, Morand, Lamy,
  Cazaux, Mercier, Dental, Deleflie, and Handschuh}]{LeFevre2014}
Le~F{\`{e}}vre C, Fraysse H, Morand H, Lamy A, Cazaux C, Mercier P, Dental C,
  Deleflie F, Handschuh D (2014) Compliance of disposal orbits with the {French
  Space Operations Act}: The good practices and the {STELA} tool. Acta
  Astronaut 94(1):234--245, \doi{10.1016/j.actaastro.2013.07.038}

\bibitem[{Liou et~al(2004)Liou, Hall, Krisko, and Opiela}]{Liou2004}
Liou JC, Hall DT, Krisko PH, Opiela JN (2004) {LEGEND} {\textendash} a
  three-dimensional {LEO}-to-{GEO} debris evolutionary model. Adv Space Res
  34(5):981--986, \doi{10.1016/j.asr.2003.02.027}

\bibitem[{{Meeus}(1998)}]{Meeus1998}
{Meeus} J (1998) {Astronomical Algorithms}. Willmann-Bell, Inc., Richmond, VA,
  USA

\bibitem[{{Milani} and {Nobili}(1987)}]{Milani1987}
{Milani} A, {Nobili} AM (1987) {Integration error over very long time spans}.
  Cel Mech 43:1--34, \doi{10.1007/BF01234550}

\bibitem[{{Morand} et~al(2013){Morand}, {Caubet}, {Deleflie}, {Daquin}, and
  {Fraysse}}]{Morand2013}
{Morand} V, {Caubet} A, {Deleflie} F, {Daquin} J, {Fraysse} H (2013) {Semi
  analytical implementation of tesseral harmonics perturbations for high
  eccentricity orbits}. In: Adv. Astronaut. Sci., Univelt, Inc., vol 150, pp
  705--722, proceedings of the AAS/AIAA Astrodynamics Specialist Conference,
  August 2013, Hilton Head, SC, USA

\bibitem[{Morbidelli(2002)}]{Morbidelli2002}
Morbidelli A (2002) Modern Celestial Mechanics. Advances in Astronomy and
  Astrophysics, Taylor \& Francis, London, United Kingdom

\bibitem[{{Moser} and {Zehnder}(2005)}]{MoserZehnder2005}
{Moser} J, {Zehnder} EJ (2005) {Notes on {D}ynamical {S}ystems}. Courant
  Lecture Notes in Mathematics, American Mathematical Society, Providence,
  Rhode Island

\bibitem[{Murray and Dermott(1999)}]{Murray1999}
Murray CD, Dermott SF (1999) Solar System Dynamics. Cambridge University Press,
  Cambridge, United Kingdom, \doi{10.1017/cbo9781139174817}

\bibitem[{Möckel(2015)}]{Moeckel2015}
Möckel M (2015) High performance propagation of large object populations in
  earth orbits. phdthesis, Technische Universität Braunschweig, Braunschweig,
  Germany

\bibitem[{{Nacozy} and {Dallas}(1977)}]{Nacozy1977}
{Nacozy} PE, {Dallas} SS (1977) {The geopotential in nonsingular orbital
  elements}. Cel Mech 15:453--466, \doi{10.1007/BF01228611}

\bibitem[{{{NASA}, {NOAA} and {US Air Force}}(1976)}]{US76}
{{NASA}, {NOAA} and {US Air Force}} (1976) {U.S. Standard Atmosphere, 1976}.
  Tech. Rep. NASA-TM-X-74335, Washington, D.C., USA

\bibitem[{Pel{\'a}ez et~al(2007)Pel{\'a}ez, Hedo, and Rodr{\'{\i}}guez~de
  Andr{\'e}s}]{Pelaez2007}
Pel{\'a}ez J, Hedo JM, Rodr{\'{\i}}guez~de Andr{\'e}s P (2007) A special
  perturbation method in orbital dynamics. Cel Mech Dyn Astr 97:131--150,
  \doi{10.1007/s10569-006-9056-3},
  \urlprefix\url{http://adsabs.harvard.edu/abs/2007CeMDA..97..131P}

\bibitem[{{Picone} et~al(2002){Picone}, {Hedin}, {Drob}, and
  {Aikin}}]{Picone2002}
{Picone} JM, {Hedin} AE, {Drob} DP, {Aikin} AC (2002) {NRLMSISE-00 empirical
  model of the atmosphere: Statistical comparisons and scientific issues}. J of
  Geophys Res: Space Physics 107(A12):15--1--16, \doi{10.1029/2002JA009430}

\bibitem[{Pines(1973)}]{Pines1973}
Pines S (1973) Uniform representation of the gravitational potential and its
  derivatives. {AIAA} J 11(11):1508--1511, \doi{10.2514/3.50619}

\bibitem[{{Radhakrishnan} and {Hindmarsh}(1993)}]{Radhakrishnan1993}
{Radhakrishnan} K, {Hindmarsh} AC (1993) {Description and Use of LSODE, the
  Livermore Solver for Ordinary Differential Equations}. Tech. Rep.
  UCRL-ID-113855, Lawrence Livermore National Laboratory

\bibitem[{Roa(2017)}]{Roa2017}
Roa J (2017) Regularization in Orbital Mechanics; Theory and Practice. De
  Gruyter, Inc., \doi{10.1515/9783110559125},
  \urlprefix\url{http://adsabs.harvard.edu/abs/2017rom..book.....R}

\bibitem[{Rossi et~al(2009)Rossi, Anselmo, Pardini, Jehn, and
  Valsecchi}]{Rossi2009}
Rossi A, Anselmo L, Pardini C, Jehn R, Valsecchi GB (2009) The new space debris
  mitigation (sdm 4.0) long term evolution code. In: Fifth European Conference
  on Space Debris, Darmstadt, Germany, ESA Special Publication, vol 672, p~90,
  \urlprefix\url{http://adsabs.harvard.edu/abs/2009ESASP.672E..90R}

\bibitem[{Sellamuthu and Sharma(2017)}]{Sellamuthu2017}
Sellamuthu H, Sharma RK (2017) Orbit theory with lunar perturbation in terms of
  kustaanheimo{\textendash}stiefel regular elements. J Guid Control Dyn
  40(5):1272--1277, \doi{10.2514/1.g002342}

\bibitem[{Sellamuthu and Sharma(2018)}]{Sellamuthu2018}
Sellamuthu H, Sharma RK (2018) Hybrid orbit propagator for small spacecraft
  using kustaanheimo{\textendash}stiefel elements. J Spacecraft Rockets
  55(5):1282--1288, \doi{10.2514/1.a34076}

\bibitem[{Setty et~al(2016)Setty, Cefola, Montenbruck, and Fiedler}]{Setty2016}
Setty S, Cefola PJ, Montenbruck O, Fiedler H (2016) Application of
  semi-analytical satellite theory orbit propagator to orbit determination for
  space object catalog maintenance. Adv Space Res 57:2218--2233,
  \doi{10.1016/j.asr.2016.02.028}

\bibitem[{{Sperling}(1961)}]{Sperling1961}
{Sperling} H (1961) {Computation of {K}eplerian {C}onic {S}ections}. J American
  Rocket Soc 31(5):660--1--661

\bibitem[{Stiefel and Scheifele(1971)}]{Stiefel1971}
Stiefel EL, Scheifele G (1971) Linear and Regular Celestial Mechanics, Die
  Grundlehren der mathematischen Wissenschaften, vol 174. Springer-Verlag,
  Berlin, Germany, \doi{10.1007/978-3-642-65027-7}

\bibitem[{Uphoff(1973)}]{Uphoff1973}
Uphoff C (1973) Numerical averaging in orbit prediction. {AIAA} J
  11(11):1512--1516, \doi{10.2514/3.50620}

\bibitem[{Vallado et~al(2006)Vallado, Crawford, Hujsak, and
  Kelso}]{Vallado2006}
Vallado DA, Crawford P, Hujsak R, Kelso TS (2006) Revisiting spacetrack report
  \#3. In: AIAA/AAS Astrodynamics Specialist Conference and Exhibit, Keystone,
  Colorado, United States, \doi{10.2514/6.2006-6753}, paper AIAA 2006-6753

\bibitem[{Williams et~al(2010)Williams, Senent, Ocampo, Ravi, and
  Davis}]{Williams2010}
Williams J, Senent JS, Ocampo C, Ravi M, Davis EC (2010) Overview and software
  architecture of the copernicus trajectory design and optimization system. In:
  4th International Conference on Astrodynamics Tools and Techniques, European
  Space Astronomy Centre (ESAC), Madrid, Spain

\end{thebibliography}

\end{document}